\documentclass[aps,prb,superscriptaddress,twocolumn,amsmath,amssymb,floatfix,eqsecnum,nofootinbib]{revtex4}
\usepackage{amsmath}  
\usepackage{epsfig}
\usepackage{bm}
\usepackage{graphicx}

\def\prl#1#2#3{Phys.\ Rev.\ Lett.\ {\bf #1}, #2 (#3)}

\def\prb#1#2#3{Phys.\ Rev.\ B {\bf #1}, #2 (#3)}

\def\prd#1#2#3{Phys.\ Rev.\ D {\bf #1}, #2 (#3)}

\def\physrev#1#2#3{Phys. Rev. {\bf #1}, #2 (#3)}
\def\npb#1#2#3{Nucl.\ Phys.\ B {\bf #1}, #2 (#3)}

\def\plb#1#2#3{Phys.\ Lett.\ B {\bf #1}, #2 (#3)}
\def\physrep#1#2#3{Phys.\ Rep.\ {\bf #1}, #2 (#3)}
\def\advphys#1#2#3{Adv.\ in Phys.\ {\bf #1}, #2 (#3)}
\def\mpla#1#2#3{Mod.\ Phys.\ Lett.\ A {\bf #1}, #2 (#3)}

\def\ijmpa#1#2#3{Int.\ J.\ Mod.\ Phys.\ A {\bf #1}, #2 (#3)}

\def\jpa#1#2#3{J.\ Phys.\ A {\bf #1}, #2 (#3)}

\def\science#1#2#3{Science {\bf #1}, #2 (#3)}

\def\cmp#1#2#3{Comm. Math. Phys. {\bf #1}, #2 (#3)}

\def\jstatphys#1#2#3{J. Stat. Phys. {\bf #1}, #2 (#3)}
\def\annphys#1#2#3{Ann. Phys (N.Y.) {\bf #1}, #2 (#3)}

\begin{document}


\title{Realizing non-Abelian statistics}

\author{Paul Fendley}

\affiliation{Department of Physics, University of Virginia,
Charlottesville, VA 22904-4714, USA}

\author{Eduardo Fradkin}

\affiliation{Department of Physics, University of Illinois at 
Urbana-Champaign, 1110 
W. Green St., Urbana, IL  61801-3080, USA
} 

\date{\today}

\begin{abstract}

We construct a series of 2+1-dimensional models whose quasiparticles
obey non-Abelian statistics.  The adiabatic transport of
quasiparticles is described by using a correspondence between the
braid matrix of the particles and the scattering matrix of
1+1-dimensional field theories.  We discuss in depth lattice and
continuum models whose braiding is that of $SO(3)$ Chern-Simons gauge
theory, including the simplest type of non-Abelian statistics,
involving just one type of quasiparticle. The
ground-state wave function of an $SO(3)$ model is related to a loop
description of the classical two-dimensional Potts model. We discuss
the transition from a topological phase to a conventionally-ordered
phase, showing in some cases there is a quantum critical point.

\end{abstract}

\pacs{PACS numbers: 
75.10.Jm, 
75.10.Hk 
} 

\maketitle

\section{Introduction}

Understanding phases with topological order has become an important
theme in condensed-matter physics. Well-understood examples of
topological fluids include the fractional quantum Hall states which
arise for electrons in two dimensions moving in large magnetic fields.
The existence of quasiparticles or quasiholes with fractional
statistics is a central and striking prediction of the theory of the
fractional quantum Hall effect and follows directly from the nature of
the electronic correlations in this quantum
fluid\cite{laughlin,arovas}. In spite of its profound conceptual
importance, only this week has there been a report of
experimental
confirmation of this startling prediction\cite{goldman}.

Non-Abelian fractional statistics are a fascinating property of some
fractional quantum Hall states \cite{moore92}. Here, the wave function
depends not only on which particles are braided, but on the order in
which they are braided: the states carry a non-Abelian (matrix)
representation of the braid group.  One of the motivations for the
current consideration of non-Abelian phases is the realization that a
physical system in a non-Abelian topological phase behaves effectively
as a universal quantum computer\cite{kitaev,freedman01,freedman03}.

The topological quantum fluids arising in the fractional quantum Hall
effect have an effective hydrodynamical description in terms of Chern-Simons
gauge theories\cite{fqh-cs,wen-zee,wen-review}.  Pure Chern-Simons is
a topological field theory, meaning that its correlators are
independent of the position of the operators and depend only on
topological invariants\cite{witten89}.  It has a vanishing
Hamiltonian; the only non-trivial properties arise from the braiding
of its Wilson and its Polyakov loops.  In this paper we are mainly
interested in theories whose {\em ground state} is topological, but
whose gapped excitations have non-Abelian statistics.

Fractional quantum Hall fluids do not have time-reversal symmetry, but
topological order occurs in models with unbroken time-reversal
invariance as well. These ``spin-liquid" phases were originally
speculated to be responsible for the unusual behavior of the ``normal
state" of high temperature superconductors\cite{pwa,krs} although, in
spite of much effort both theoretical and experimental, there is yet
no solid evidence for any spin-liquid state. Nevertheless, as a
consequence of much (theoretical) effort, we know that topological
order occurs in the ground states in certain gapped systems with
time-reversal invariance and reasonably-local
interactions\cite{kitaev}, for example in quantum dimer models on
non-bipartite lattices\cite{RK,moessner-sondhi}. Morally, these
topological phases are equivalent (in the sense of asymptotic low
energy theories) to deconfined phases of an effective gauge
theory. The excitations of these time-reversal invariant phases do
exhibit electron
fractionalization\cite{kivelson87,read-chak,read-sachdev,senthil-fisher},
but the statistics is Abelian.

Non-Abelian topological phases are even harder to come by.  Such
phases with broken time-reversal symmetry do occur in the fermionic
Pfaffian (a.k.a.\ Moore-Read) wave function \cite{moore92} for the
$\nu =5/2$ fractional quantum Hall effect, and in the $\nu=1$ bosonic
Pfaffian state\cite{read-rezayi,FNS,ardonne99}.  The former occurs in
models of $p$-wave superconductors as well \cite{ivanov}. Field
theories with such non-abelian statistics also have been found
\cite{alford,Lo,bais}. In a number of these cases, the long-distance
physics can be described by a Chern-Simons gauge theory, which breaks
time-reversal symmetry.  Non-abelian topological phases, however,
occur in time-reversal-invariant systems as
well\cite{freedman01,freedman03}. The resulting effective Chern-Simons
theory is doubled to restore the time-reversal symmetry. To give the
theory a gap while keeping the topological theory as its ground state,
one can include the electric-field part of the Maxwell term $\int d^2
x\; \vec{E}\cdot \vec{E}$ in the Hamiltonian
\cite{witten92,grignani,ardonne}. Hence in these topological phases,
the ground-state wave function is a superposition of configurations of
Wilson loops in two-dimensional space, while the world-lines of the
excitations correspond to Polyakov loops in the 2+1-dimensional field
theory \cite{freedman03}. For example, the lattice models discussed in
detail in Refs.\ \onlinecite{freedman-nayak-shtengel-03,freedman04}
have a continuum description in terms of doubled $SU(2)_k$
Chern-Simons gauge theory. The resulting configuration space is
naturally associated with the Temperley-Lieb algebra\cite{TL}.

A natural description of the configuration space of models in a
topological phase is in terms of loops
\cite{kitaev,freedman01,freedman03,freedman-nayak-shtengel-03,freedman04}.
This holds in both Abelian and non-Abelian cases.
Precisely, each basis state in the Hilbert space of the quantum theory
is a loop configuration in two dimensions. Quantum dimer models and
their generalizations \cite{ardonne,chamon} can also be viewed
as quantum loop gases.

In this paper we reexamine the problem of non-Abelian topological
phases by starting with the statistics we wish to have, and working
backward to construct a model which exhibits them. We thus first give
an algebraic way of characterizing the braiding in both $SU(2)_k$ and
$SO(3)_k$ Chern-Simons theories. We show that such a braid matrix of
a 2+1-dimensional theory is a limit of the $S$-matrix of an
associated relativistic 1+1 dimensional model, and give an intuitive
argument as to why this is so.  We then show how to
explicitly construct quantum two-dimensional models with these braid
relations by utilizing the structure of the factorizable $S$-matrices
of integrable 1+1-dimensional models.

Specifically, we embed the
1+1-dimensional model in two-dimensional Euclidean space, and find a
Rokhsar-Kivelson-type quantum Hamiltonian\cite{RK} acting on this
two-dimensional space whose ground state has the
properties expected of a model with non-Abelian statistics. 
In both cases we discuss in detail, the Hilbert space is that of a
loop gas: in the $SU(2)_k$ case, the loops are self-avoiding and
mutually-avoiding\cite{freedman04}, while in the $SO(3)_k$ case, the
loops intersect. The latter are thus more akin to nets than
loops\cite{levin03}. Both these loop gases are associated with
well-known two-dimensional classical statistical mechanical models:
in the $SU(2)_k$ case, this is the known as the $O(n)$ lattice model
with $n=2\cos(\pi/(k+2))$,
while in the $SO(3)_k$ case, this is the $Q$-state Potts model with $Q=4\cos^2(\pi/(k+2))$. The loop
expansion of the former is well known\cite{nienhuis}, but the one we
utilize for the Potts model does not seem to have been discussed before.

Having an explicit lattice construction of the
states enables us to construct (reasonably) local quantum lattice
models with these ground state wave functions. By studying the
statistical properties of the absolute value squared of these wave
functions, we can
investigate the correlations described by these quantum states, and
determine if they describe quantum critical points or massive
(topological) phases. Both here and in the $SU(2)$
case\cite{freedman-nayak-shtengel-03}, the result depends on the level $k$.

The paper is organized as follows. In Section \ref{sec:braids} we
describe the algebraic approach to non-Abelian statistics in
both the $SU(2)_k$ case and $SO(3)_k$ cases.  In Section
\ref{sec:qlg} we discuss quantum loop gases and their relation to the
$S$ matrix of 1+1-dimensional integrable field theories. In Section
\ref{sec:examples} we give an explicit construction of these $S$-matrices
for $SO(3)$ and $SU(2)$ braiding. In Section \ref{sec:field} we
discuss the corresponding 1+1-dimensional field theories. In Section
\ref{sec:lattice} we discuss lattice models whose ground states are
precisely the loop wave functions, both $SU(2)_k$ and $SO(3)_k$, with
the desired braiding properties. We give
a set of specific criteria that 2+1-dimensional quantum Hamiltonian
ought to satisfy and give an explicit construction.  In Section
\ref{sec:phase} we discuss under what circumstances these
wave functions are topological and when do they describe quantum
critical systems. In Appendix \ref{app:RSOS} we give a summary of the
Landau-Ginzburg description of the 1+1-dimensional theories whose
$S$ matrix we use.

\section{Braids and Algebras}
\label{sec:braids}

Particle statistics, of course, are the effect on the wave function
when particles are adiabatically transported around each other a large
distance away.  This picture follows from the familiar concept of
adiabatic particle transport, developed in detail in the context of
the Laughlin states of the fractional quantum Hall effect where
it follows from the Berry phase accumulated during an adiabatic
evolution of the state with two quasiparticles\cite{arovas}.

Adiabatic particle transport can be represented pictorially
by drawing the world lines of the particles, which are the paths they
trace out in space-time.  Since the particles stay far apart, we need
only study paths which do not cross.  (Non-trivial braid statistics
always require the assumption that particles have a hard-core short
distance repulsion).  The world lines of the particles therefore {\em
braid} around each other.  Formally, the set of all possible braidings
is a group, acting on the space of states of the system.  Different
types of statistics correspond to different representations of the
braid group. In this paper we consider two spatial dimensions, where
both Abelian and non-Abelian statistics are possible. A system whose
quasiparticles are associated with a non-Abelian representation of the
braid group has a degenerate set of states of quasiparticles at fixed
positions ${x_1}, {x_2}, \ldots, {x_N}$. Call this space of states
$V(N)$. The states in $V(N)$ are locally indistinguishable but differ
topologically. When a quasiparticle is taken around another, states in
this degenerate subspace are rotated into each other.  Since this is
quantum mechanics, it can take a state to a linear combination of
other states: adiabatic particle transport can entangle the states.

Although we will not require its use, it is useful to note that
non-Abelian statistics can be implemented in a local field theory by
assigning to each quasiparticle a charge and a flux under a
non-Abelian gauge group.  Within this picture, a quasiparticle carries
a representation $V$ (i.e. a `charge') of the gauge group $G$ and a
flux $g\in G$. When quasiparticle $1$ is taken around quasiparticle 2,
its internal state ${\psi_1}\in V_1$ is rotated to ${g_2}\psi_1$ by
the action of the flux ${g_2}\in G$ which is associated with
quasiparticle 2.  It is worth keeping in mind that this picture is not
gauge invariant, and there is, in fact, no local degree of freedom
associated with each quasiparticle (since a gauge transformation can
change $\psi\in V$ into any other $\psi'\in V$ and can change the flux
$g$ into any other element of its conjugacy class
$hgh^{-1}$).

Studying the statistics of a 2+1-dimensional system can effectively be
reduced to a two-dimensional problem. We project the
world lines onto the plane (ignoring boundary conditions), and call them {\em strands}. When there are $N$ particles, we
have $N$ strands.
A braiding in 2+1 dimensions results in the
crossing of two strands in the two-dimensional picture. 
In this projection, there are overcrossings and
undercrossings. As long as we are only interested in the statistics
of the particles, the other details of the projection are not
particularly important: we can move the strands around at will as long
as we do not remove crossings or create new ones.

It is useful to think of this collection of strands in the plane in a
1+1-dimensional fashion. The degenerate
states of the 2+1-dimensional system correspond to a set of degenerate
multi-particle states in a one-dimensional quantum system. In this 1+1-dimensional
quantum system, each strand is the world line
of a real local degree of freedom. Heuristically, this is the
``gauge-fixed'' version of the model.
Consider a configuration at $t=-\infty$
({\it i.e.\/}  before any of the particles have been braided) where all the
particles are very far from each other ({\it i.e.\/} all at spatial
infinity). We can thus consider these particles to all be on a
circle. This circle is our one-dimensional space. We can construct the
full configuration by a sequence of adiabatic braidings as the
particles move inward.  Since we are free to move around the strands
as long as we do not add or remove crossings, we can then take all the
particles at $t=\infty$ to lie on a circle as well. Thus we can
project all the original 2+1-dimensional world-lines onto a
two-dimensional annulus. We can then view the angular direction of the
annulus as one-dimensional space, and the radial direction as
(Euclidean) time.  Thus, each configuration in the plane can be
regarded as an adiabatic evolution in an equivalent 1+1-dimensional
Euclidean field theory in which each strand (or particle) belongs to a
given Hilbert space associated with the species.

\begin{figure}[h!] 
\begin{center} 
\includegraphics[width= .4\textwidth]{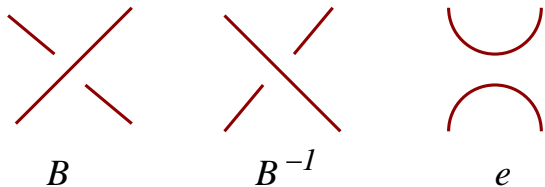} 
\caption{braid, reverse braid, and Temperley-Lieb generator} 
\label{fig:braid} 
\end{center} 
\end{figure}

\begin{figure}[h!] 
\begin{center} 
\includegraphics[width= .28\textwidth]{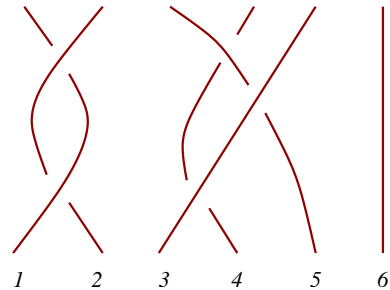} 
\caption{A typical braiding involving six particles}
\label{fig:ybraids}
\end{center} 
\end{figure}

\begin{figure}[h!] 
\begin{center} 
\includegraphics[width= .4\textwidth]{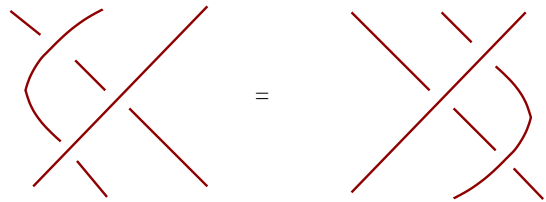} 
\caption{Consistency relation for braiding} 
\label{fig:ybe} 
\end{center} 
\end{figure}

To make this more precise, let us consider the $N$-particle space of
states $V(N)$ on which the braid group acts. For now we let this space
be the tensor product of $N$ copies of the single-particle space of
states: $V(N)=V^{\otimes N} \equiv V_1\otimes V_2 \otimes\dots \otimes
V_N $. (Later, we will see that the actual space of states is a
subspace of $V^{\otimes N}$.) In the 1+1 dimensional picture, we can
think of $V^{\otimes N}$ as the space of particles on a circle.  The
elements of the braid group corresponding to overcrossings and
undercrossings are denoted $B_i$ and $B_i^{-1}$ and are displayed in
Fig.\ \ref{fig:braid}.
The subscript $i$ means that $B_i$ is
describing the crossing between the $i$th and the $(i+1)$th particles,
and so acts non-trivially on the space $V_i \otimes V_{i+1}$, and with
the identity on the other spaces $V_j$, with $j\neq i,i+1$.
For example, the braids in Fig. \ref{fig:ybraids}
are described algebraically as $=B_3^{-1} B_4 B_3 B_1 B_1$.
The braid-group generators $B_i$ must satisfy the relations
\begin{eqnarray}
\nonumber
B_i B_{i+1} B_i &=& B_{i+1} B_i B_{i+1},\\
B_i B_j &=& B_j B_i \qquad\qquad |i-j| \ge 2.
\label{braid}
\end{eqnarray}
These relate configurations which are topologically identical,
as can easily be seen from Fig. \ref{fig:ybe}.

If the matrices $B_i$ are diagonal, then the statistics are
Abelian. For bosons the $B_i$ matrices are all the identity; 
for anyons their
entries are phases. In this paper we are interested in non-Abelian
representations of the braid group, so that particles obey non-Abelian
statistics: the wave function changes form depending on the order in
which the particles are braided.  One can give explicit matrix
representations of the braid group. However, it is usually much
convenient to study the algebra of the matrices involved.  In the
cases of interest here, the statistics of the particles can be
obtained directly from the algebraic relations the matrices obey,
without need for their explicit representation.

\subsection{The $SU(2)$ theory}

A famous one-parameter set of non-Abelian representations of the braid
group arises from
utilizing the Temperley-Lieb algebra \cite{TL}.  These representations
give rise to the Jones polynomial in knot theory
\cite{Jones,Kauffman,ADW}, and correspond to the braiding of Wilson
and Polyakov loops in $SU(2)$ Chern-Simons theory.

\begin{figure}[h] 
\begin{center} 
\includegraphics[width= .48\textwidth]{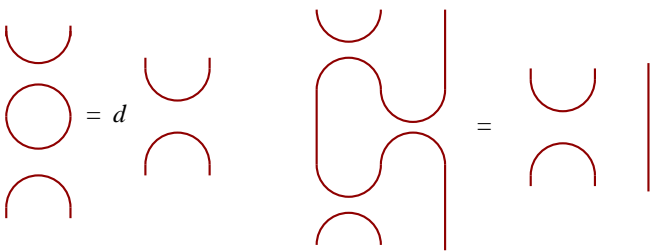} 
\caption{The Temperley-Lieb algebra} 
\label{fig:TLalg} 
\end{center} 
\end{figure}

The Temperley-Lieb algebra originally arose
as a way of relating the Potts models to the six-vertex model.  The
transfer matrices of these two models (and a number of other models)
can be written in terms of different representations of this
algebra. This means that any properties of the model which can be
computed from purely algebraic considerations will be the same for any
such model.
A generator of the Temperley-Lieb algebra $e_i$ acts non-trivially on
the $i$th and $(i+1)$th particles; a useful pictorial representation is
given in Fig.\ \ref{fig:braid}. The algebra is \cite{TL}
\begin{eqnarray}
e_i^2 &=& d e_i,\nonumber\\ e_i\, e_{i\pm 1}\, e_i &=&
e_i,\nonumber\\  e_i\,e_j&=&e_j\,e_i \quad (|j-i|\ge 2).
\label{TLalg}
\end{eqnarray} %
where $d$ is a parameter. These algebraic relations are drawn in 
Fig.\ \ref{fig:TLalg}.
From the picture, one can see that 
$d$ can be thought of as the weight of a closed loop.

\begin{figure}[h!] 
\begin{center} 
\includegraphics[width= .4\textwidth]{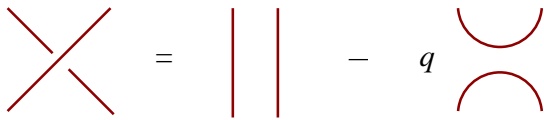} 
\caption{The braid in terms of the Temperley-Lieb generator} 
\label{fig:BIE} 
\end{center} 
\end{figure}

Representations of the braid group can be found from representations
of the Temperley-Lieb algebra by letting
\begin{equation}
B_i = I  -q e_i,
\label{Be}
\end{equation}
where $I$ is the identity.
This is illustrated in Fig. \ref{fig:BIE}.
The $B_i$ defined in this fashion obey the braid-group relation
(\ref{braid}) when
$$d = q+q^{-1}.$$
It is also easy to check that
$$B_i^{-1} = I - q^{-1} e_i.$$ Note that in writing the braid group in
this fashion, we have resolved the crossings $B$ and $B^{-1}$ in terms
of strands which no longer intersect.

These braid relations are those of Wilson
loops in $SU(2)_k$ Chern-Simons theory when \cite{witten89}
$$d = 2 \cos\left(\frac{\pi}{k+2}\right)$$
or equivalently $q = e^{i\pi/(k+2)}$. 
The integer $k$ is the coefficient of the Chern-Simons term in the
gauge theory, and is known as the level. It must be an integer to
ensure gauge invariance of the Chern-Simons gauge theory \cite{deserjackiw}.

The configuration space of Wilson loops on the plane is a very large
space and arbitrary superpositions of these states do not obviously
describe a topological ground state. It turns out that in order to
enforce the condition that states be purely topological, these ground
states must satisfy additional properties which can be enforced by
means of a suitable projection operator. This is known as the
Jones-Wenzl projector, which acts on $k+1$ strands without
crossings. For details on its application in this context, see
Refs. \onlinecite{freedman01,freedman03}. We will give an algebraic
description of this below. By construction, states obtained by means
of this projector are topological and do not support local low energy
degrees of freedom. Conversely, unprojected states can describe low
energy, even massless, degrees of freedom, and are unphysical states
in a topological gauge theory such as Chern-Simons. However,
unprojected states may describe degrees of freedom associated with
quantum critical points.

\subsection{The $SO(3)$ theory}
\label{subsec:bmw}

Here we discuss another one-parameter set of non-Abelian
representations of the braid group. These describe the braiding in
$SO(3)$ Chern-Simons theory, instead
of $SU(2)$. The algebras are of course the same; the key distinction
is that Wilson and Polyakov loops in $SO(3)$ Chern-Simons occur in
only integer-spin representations. The corresponding representation
of the braid group is given in terms of the $SO(3)$
Birman-Murakami-Wenzl (BMW) algebra \cite{BMW}, defined below.  This
algebra has two non-trivial generators $X_j$ and $E_j$ acting on
adjacent strands.

\begin{figure}[h!] 
\begin{center} 
\includegraphics[width= .3\textwidth]{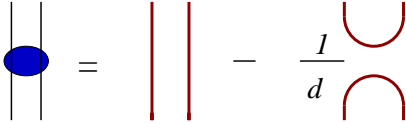} 
\caption{Projecting onto the spin-1 representation} 
\label{fig:proj} 
\end{center} 
\end{figure}

The $SO(3)$ braid relations can be found by ``fusing'' together
two strands obeying the Temperley-Lieb algebra: the $SO(3)$ BMW generators
$X_j$ and $E_j$ can be written in terms of the Temperley-Lieb
generators $e_i$.  Heuristically, the idea is to exploit the fact that
a spin-1 representation of $SO(3)$ can be found from the tensor
product of two spin-$1/2$ representations of $SU(2)$. This statement
is still true in the ``quantum-group'' algebra $U_q(sl_2)$, which is a
one-parameter deformation of the ordinary Lie algebra $sl_2$. One can
define an action of $U_q(sl_2)$ on the space of states $V^{\otimes N}$
which commutes with the $e_i$; see Ref. \onlinecite{slingerland} for an extensive
discussion of quantum groups in the context of non-Abelian statistics
($q$ there is $q^2$ here).  In particular, to relate the two algebras,
first note that $e_i/d$ is a projector. In $U_q(sl_2)$ language, this
projects onto the trivial spin-0 representation.  The projector 
onto the spin-1 representation is therefore
\begin{equation}
P_i = I - \frac{1}{d}\, e_i,
\label{proj}
\end{equation}
so that $P_i e_i =0$.
The single-particle space of states $W_j$ in the $so(3)$ BMW algebra
is comprised of two ``fused'' Temperley-Lieb strands, projected onto
the spin-1 representation. In an equation, $W_j = P_{2j-1}
[V_{2j-1}\otimes V_{2j}]$.  Pictorially, just think of each strand in
the $so(3)$ theory as the left-hand-side of Fig.\ \ref{fig:proj}.

\begin{figure}[h!] 
\begin{center} 
\includegraphics[width= .45\textwidth]{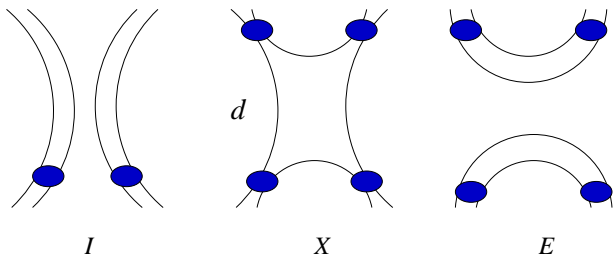} 
\caption{The generators of the $SO(3)$ BMW algebra} 
\label{fig:IXE} 
\end{center} 
\end{figure}

With this identification, the $so(3)$ BMW algebra follows from the
Temperley-Lieb algebra. 
Since lines never cross in the latter, they cannot cross in the former either.
When two of the fused strands come near each
other, there are now three possibilities for what happens, which we display in
Fig.\ \ref{fig:IXE}. 
{}From the pictures, we read off that
\begin{eqnarray}
E_{j} &=& {P}_{2j-1}\, {P}_{2j+1}\, e_{2j}\,
e_{2j-1}\, e_{2j+1}\, e_{2j}\, {P}_{2j-1}\, {
P}_{2j+1},\cr\cr X_{j} &=& d\ {P}_{2j-1}\, {P}_{2j+1}\,
e_{2j}\, {P}_{2j-1}\, {P}_{2j+1}. \label{EXTL}
\end{eqnarray}
These generators act on the two-particle states in $W_j\otimes
W_{j+1}$. It is straightforward to verify using (\ref{TLalg}) that they
obey the $o(3)$ BMW algebra. We have
\begin{eqnarray}
\nonumber
(E_i)^2 &=& (Q-1) E_i,\\
\nonumber
(X_i)^2 &=& (Q-2) X_i +  E_i.\\
E_iX_i =X_i E_i &=& (Q-1) E_i.
\label{EXalg}
\end{eqnarray}
where the parameter $Q\equiv d^2$. 

Relations involving generators on adjacent sites
({\it e.g.\/} $E_iE_{i+1}E_i=E_i$) are straightforward to work out using
the Temperley-Lieb algebra; they can be found for example in
Ref.\ \onlinecite{fendleyread}. Most become fairly obvious after drawing
the appropriate picture.
The relations involving only the $E_i$  are
those of the Temperley-Lieb algebra \eqref{TLalg}, but with $d$
replaced here by $Q-1$, so that closed isolated loops of ``spin-1'' particles
get a weight $Q-1=d^2-1 = 1+q^2+q^{-2}$.
This factor of $d^2-1$ is obvious from the pictures;
the $d^2$ comes from the $d$ from each loop,
and $1$ must be subtracted because of the projection onto spin-1 states
(projecting out the singlet).

The reason we have done all this work is to give us another
representation of the braid group. Namely, defining
\begin{eqnarray}
B^{SO(3)}_j&=&  q^2 {\cal I} -  X_j + q^{-2}E_j,
\label{braidBMW}
\end{eqnarray}
it is straightforward to check using the $so(3)$ BMW relations 
that the $B_j$ satisfy the braid-group relations
(\ref{braid}). ${\cal I}$ is the identity on 
the projected Hilbert space
$W^{\otimes N}$; on 
$V^{\otimes N}$, we have 
${\cal I}=P_{2j-1}P_{2j+1}$ (see Fig.\ref{fig:IXE}).
One can also check that 
$$(B^{SO(3)}_j)^{-1}= q^{-2} {\cal I} - X_j + q^2E_j.$$
Particles with braiding given by $B^{SO(3)}_j$ arise from $SO(3)$
Chern-Simons theory. This follows from our construction: we basically
have restricted the particles to be associated with integer-spin
representations of $U_q(sl_2)$; this is precisely what one does to go
from $SU(2)$ to $SO(3)$. 

\subsection{The Jones-Wenzl projector}
\label{subsec:jones}

The Jones-Wenzl projector is simply expressed in terms 
the projector ${\cal P}^{(s)}$ onto the representation of
spin $s$ of $U_q(sl_2)$: for a given $k$ it is simply ${\cal
 P}^{([k+1]/2)}_j=0$ for all $j$. 
This projector involves
$k+1$ strands, so this amounts to being able to replace the identity
acting on $k+1$ strands with a linear combination of other
Temperley-Lieb or BMW elements. The necessity of this projection
is also apparent from the representation theory of $U_q(sl_2)$: 
when $q^{k+2}=-1$, the 
spin-$(k+1)/2$ representation is reducible but is indecomposable (it
cannot be written as a direct sum of irreducible
representations). Performing the projection avoids all sorts of
complications such as zero-norm states.

We have already seen one example of such a projector.  The projector
$P_i = I - e_i/d$ is the projector onto the spin-1 representation of
the quantum-group algebra, so ${\cal P}^{(1)}_j =P_j$.  When $k=1\
(d=1)$, the Jones-Wenzl projector is simply $P_j$: any spin-1
combination of strands is projected out. Therefore the $SO(3)$ theory at
$k=1$ is trivial.

A case of great interest is $SO(3)_3$, the ``Lee-Yang'' model. This is
the simplest model of non-abelian statistics, because there is only
one type of non-trivial braiding.  The naming arises because the
braiding relation for the particle in this model is the same as the
fusion rule associated with the Lee-Yang conformal field
theory\cite{cardy}.  We have for any $k$ in both the $SU(2)$ and
$SO(3)$ models
\begin{equation}
{\cal P}^{(2)}_j = P_{2j-1} P_{2j+1} - \frac{1}{d^2-2} X_j
+ \frac{1}{(d^2-2)(d^2-1)} E_j
\label{P2}
\end{equation}
When $k=3$, the Jones-Wenzl projector sets ${\cal P}^{(2)}=0$. In the
$SU(2)_3$ theory, this is a relation involving four strands, while in the
$SO(3)_3$ theory, this relates two of the fused strands. This means
that in $SO(3)_3$, imposing the Jones-Wenzl projector allows us to
replace any appearance of a generator $X_j$ in favor
of ${\cal I}$ and the $E_j$: we have
$$X_j = (Q-2) ({\cal I} + E_j)\qquad\qquad\hbox{for }SO(3)_3.$$
Plugging this back into the braid relation
(\ref{braidBMW}) and remembering that
$Q-2=q^2+q^{-2}$ gives
\begin{eqnarray}
B^{SO(3)_3}_j&=&  -q^2 {\cal I} - q^{-2}E_j.
\label{braidBMW3}
\end{eqnarray}

Note that for this value $k=3$ (and this value alone), the generators
$E_j$ obey the same Temperley-Lieb algebra as the $e_i$, because
$d=2\cos(\pi/5)=d^2-1$. Moreover, $q^5=-1$ here, so the braid
generator (\ref{braidBMW3}) is equivalent to that in (\ref{Be}). We
thus are led to an intriguing result: the $O(3)_3$ theory is almost
identical to that of $SU(2)_3$. There is one important difference: in
$SO(3)_3$, we have already imposed the Jones-Wenzl projector, while in
$SU(2)_3$ this still needs to be imposed. Thus (locally) $SO(3)_3$
with the Jones-Wenzl projection imposed is equivalent to $SU(2)_3$
{\em without} it imposed.

\section{Quantum loop gases and the $S$ matrix}
\label{sec:qlg}

In the preceding section we discussed some of the marvelous properties
of particles with non-Abelian statistics. Now we discuss
how to associate the $S$ matrix of a
relativistic 1+1-dimensional field theory to the type of braiding
discussed above. We argue that knowing the $S$
matrix in this 1+1-dimensional theory allows us to find a quantum
loop gas in {\em two space dimensions} where the quasiparticles
should have these braid relations.

A natural place to look for models with non-Abelian statistics is in
{\em quantum loop gases}. The reason is that 
if one projects the world lines onto the (spatial) plane, one obtains
loops: strands cannot end.
It is also natural from the point-of-view of field theory:
in pure Chern-Simons theory the only gauge-invariant degrees of
freedom are loops. In a 2+1-dimensional picture, this means it is a
good idea to look for a system where the low-energy degrees of freedom
are loops in the plane, a quantum loop gas.  
In a number of cases it has been argued that quantum loop gases turn
into gauge theories with Chern-Simons theories in the continuum
\cite{freedman01,freedman03,levin03,ardonne}.  The excitations can be
non-Abelian in a {\em topological phase}, where the ground state
contains a superposition of Wilson loops (loops in the spatial
plane). The excited states are Polyakov loops, loops which extend in
the time direction.  The quasiparticles have non-Abelian statistics
when the gauge group is non-Abelian with level $k>1$, Ref. \onlinecite{witten89}.

To understand these loop gases, it is best to first focus on the
properties of the ground state.  The types of ground states we are
interested in are liquid states, where all local order parameters have
vanishing expectation values. Such a ground state is a superposition
of different loop configurations. 
The basis states $|s\rangle$ 
of the Hilbert space can all be
described by some configuration of loops in two dimensions. In these models,
the wave function $\Psi$ of this ground state can be written in the form
\begin{equation}
\langle s|\Psi \rangle = 
\frac{e^{-{\cal S}(s)}}{Z}
\label{spsi}
\end{equation}
where ${\cal S}(s)$ turns out to be the
action of the {\em classical} two-dimensional loop model for the
configuration corresponding to $s$. $Z$ is the usual two-dimensional
partition function with weight $|\langle s|\Psi \rangle|^2$, which is 
the functional integral over all configurations $s$
with weight $e^{-{\cal S}(s)  -{\cal S}^*(s)}$. However, just because the loops can be non-local does not
mean the action needs to have long-range interactions.  Recall, for example, that
one can describe all the configurations in the classical two-dimensional Ising model in terms of
closed loops of arbitrary length, the domain walls. Nevertheless, the
interactions are still local.

Since we are identifying the loops with the quasiparticle world
lines, we need to find a quantum loop gas whose ground-state
wave function satisfies the appropriate braiding properties. To make
this notion precise, let us examine the classical loop gas ({\it i.e.\/} the
one with action ${\cal S}(s)$) corresponding to the ground state.  Now view
this two-dimensional loop gas as a 1+1-dimensional quantum
system. The loops then can be thought of as world lines of particles
in this 1+1-dimensional system. Their wave function is then a vector
in the space $V^{\otimes N}$, just like before. The braid generators
act in the same way as well.

To summarize the arguments so far: we project the world lines of the
particles in 2+1 dimensions down onto the plane, so that they form
loops. A two-dimensional quantum system possessing such particles is a
loop gas, where the degrees of freedom are loops in the plane.  The
ground-state wave function of the quantum system is the expressed in
terms of the action ${\cal S}(s)$ of the corresponding classical
two-dimensional loop gas. Finally, we then identify these loops as the
world lines of particles in the corresponding 1+1-dimensional
problem. The upshot is that by restricting ourself to considering the
ground state, we have reduced a 2+1-dimensional problem to a
1+1-dimensional one.
 Theories for which this construction holds are inherently {\em holographic} in that the 
degrees of freedom can be naturally projected to a boundary.

In the 1+1-dimensional theory, when two particle world lines cross,
the $S$ matrix plays the role of the braid matrix.
In other words, it describes what happens to the wave function when the
paths of two particles cross.
Consider the wave function describing two particles of momentum and
position $p_i,x_i$ and $p_{i+1},x_{i+1}$ respectively. The $S$ matrix
is a matching condition on the wave functions for $x_i\ll x_{i+1}$ and $x_i
\gg x_{i+1}$. As before, the wave function is a vector in $V^{\otimes N}$.
The two-particle $S$ matrix for scattering particle
$i$ from particle $i+1$ acts non-trivially in $V_i\otimes
V_{i+1}$:
$$\psi^{}_{V_i\otimes V_{i+1}} ({x_i \gg x_{i+1}}) = 
S_i (p_i,p_{i+1}) \psi^{}_{V_i\otimes V_{i+1}} (x_i \ll x_{i+1}) . $$ Our
theories are rotationally invariant in two-dimensional space, so the
corresponding 1+1-dimensional theory is Lorentz invariant. This
means that the $S$ matrix depends only on the relative rapidity
$\theta$: defining $p_i=m\sinh\theta_i$ and
$p_{i+1}=m\sinh\theta_{i+1}$, we have $\theta\equiv
\theta_1-\theta_2$.
We note that the $S$ matrix here should not 
be confused with what is usually called
the modular $S$ matrix, which governs the braidings in
Chern-Simons theory and in conformal field theory.

This correspondence between the braid group and the $S$ matrix has
long been known, in the context of knot theory \cite{ADW}.
Representations of the braid group (and the resulting knot invariants)
can be found by taking a special limit of solutions of the Yang-Baxter
equation \cite{ADW}. In physics, the Yang-Baxter equation arises in
integrable lattice models and field theories. In integrable lattice models, the
Boltzmann weights must satisfy Yang-Baxter; in the former, the $S$ matrices of
the particles do! Note, moreover, that the braid matrices and the $S$
matrices are acting in the same space $V^{\otimes N}$. So our
arguments indicate that {\em the braid matrices of the 2+1-dimensional
theory are a {\em limit} of the $S$ matrices of the corresponding 1+1-dimensional theory}.

It is not difficult to find in which limit this holds.
The Yang-Baxter equation for the $S$ matrix 
arises from requiring that the three-body $S$ matrix factorizes into a
product of two-body ones. Since there are two different ways of
factorizing, for consistency one must have
\begin{eqnarray}
\nonumber
S_i(\theta_1-\theta_2) S_{i+1}(\theta_1-\theta_3) S_i (\theta_2-\theta_3)\qquad&&\\ 
=  S_{i+1}(\theta_2-\theta_3) S_i(\theta_1-\theta_3) 
S_{i+1}(\theta_1-\theta_2).&&
\label{ybe}
\end{eqnarray}
The connection to the braid group is now obvious:
$S(0)$ and $S(\infty)$ obey the braid group relation (\ref{braid}).
In most known cases (and in the cases of interest here)
$S(0) \propto I$, and $S(-\theta)S(\theta) =I$. Thus we have
\begin{eqnarray}
\nonumber 
B &=& \lim_{\theta\to\infty} \widetilde{S}(\theta)\\
B^{-1} &=& \lim_{\theta\to\infty} \widetilde{S}(-\theta)
\label{BS}
\end{eqnarray}
The matrix $\widetilde{S} = e^{i\delta(\theta)} e^{i\theta A} S
e^{i\theta B}$, where $\delta(\theta)$ is a function of the rapidity $\theta$, and $A$ and $B$
are diagonal $\theta$-independent matrices. These factors arise in
general to ensure that $S$ has the correct properties under crossing
symmetry and unitarity. Obviously, we need to remove the oscillating
factors as $\theta\to\infty$ to have a well defined limit.  Both $S(\theta)$ and
the modified matrix $\widetilde{S}(\theta)$ satisfy the Yang-Baxter
equation \eqref{ybe}.

The limit $\theta\to\infty$ in Eq.\ \eqref{BS}
also makes sense at an intuitive level. In order for the
$S$ matrix to be that of a loop gas, one should be in a limit where
the mass $m$ of the particles is small: otherwise, the loops would be
high in energy and not dominate the partition function. When the
particle mass is small, one can create particles of any rapidity
$\theta_i$, and so the rapidity difference
$\theta=\theta_i-\theta_{i+1}$ will typically be large.

To conclude this section, we note that there are two important
additional steps to take in constructing a quantum loop gas having
quasiparticles with non-Abelian statistics. The first is to find a
Hamiltonian which has the ground-state wave function of Eq.\
\eqref{spsi}. This can usually be done by a trick utilized by Rokhsar
and Kivelson \cite{RK}.  This trick is useful for any field theory
with an explicit real action \cite{arovas91,ardonne}. For
lattice models, there can be complications, because one cannot always
construct a Hamiltonian which is ergodic in the Hilbert
space. Nevertheless, in many cases of interest this procedure has been
successful.

The second additional step is to make sure that the excited states
have braid relations which are those of the loops in the ground
state. One way of doing this is to have a Hamiltonian so that the
excited states are {\em defects} in the configuration space of loops. That is, a particle and an antiparticle over
the ground state are connected by a strand. Thus when they are moved
around each other, the non-locality due to the strand results in the
braid relations described above. As is well known, this construction works in the Abelian case.

\section{The braid matrices}
\label{sec:examples}

In this section we give explicit expressions for the $S$ matrices and
braid matrices associated with the Temperley-Lieb and BMW algebras
described in section \ref{sec:braids}.
This will enable us in the next section to identify the
two-dimensional classical field theories associated with these 1+1-dimensional
quantum theories, so that we can construct quantum
loop gases with the desired braiding.

In the $SU(2)$ case, the correspondence given in Eq.\ \eqref{BS} means that
we need to look for an $S$ matrix which at infinite rapidities is of
the form of Eq.\eqref{Be}. Such an $S$ matrix has been known for quite
some time\cite{su2-S-matrix}. It is straightforward to check that
\begin{equation}
\widetilde{S}_i(\theta) = I - \frac{e^{\lambda \theta} - e^{-\lambda \theta}}
{q^{-1} e^{\lambda\theta} - q e^{-\lambda\theta}}\, e_i
\label{STL}
\end{equation}
obeys the Yang-Baxter equation (\ref{ybe}) for any value of the parameter
$\lambda$,
as long as $e_i$ satisfies the Temperley-Lieb algebra,
Eq.\eqref{TLalg}.

A number of related models have $S$ matrices which can be written in
the form of Eq.\eqref{STL}.  The most famous is the sine-Gordon
model.  There are two different particles, the
soliton (labeled $+$) and antisoliton (labeled $-$), forming the
spin-1/2 representation of the $U_q(sl_2)$ symmetry of
the model. The single-particle space of states $V_i$ is
two-dimensional, so that $e_i$ (which acts on $V_i\otimes V_{i+1}$) is
a four-by-four matrix in this representation of the Temperley-Lieb
algebra. Labeling the rows and columns in the order $++$, $+-$, $-+$, $--$
gives
\begin{equation}
e^{6v}_i =
\begin{pmatrix}
0&0&0&0\cr
0&q&1&0\cr
0&1&q^{-1}&0\cr
0&0&0&0
\end{pmatrix}
\label{eSG}
\end{equation}
We have labeled this with a $6v$ because the
the Boltzmann weights of the six-vertex model can also be expressed in
terms of these $e_i$.

This particular representation $e^{6v}$ of the Temperley-Lieb algebra,
however, does not result in the braid matrices of the 2+1-dimensional
theory, as it does not respect the Jones-Wenzl projector. Namely,
consider $k+2$ strands in a row, {\it i.e.\/} the space $V_1\otimes
V_2\otimes\dots\otimes V_{k+2}$.  Any $k+1$ strands in a row must obey
the Jones-Wenzl projector: we must restrict the space of states so
that ${\cal P}_1^{([k+1]/2)}={\cal P}_2^{([k+1]/2)}=0$ (the former
acts non-trivially on the first $k+1$ strands, the latter on the
strands $2\dots k+2$). Now braid the $k+2$th particle with the
$k+1$st; the resulting configuration need no longer satisfy ${\cal
P}^{([k+1]/2)}_1=0$. For example, consider three strands in the $k=1$
case, where configuration $(-+-)$ in $V_1\otimes V_2\otimes V_3$ is
part of the projected Hilbert space. If we braid the last two
particles, the off-diagonal terms in $e^{6v}$ result in a non-zero
amplitude for the final state $(--+)$. The latter state is projected
out of the Hilbert space, since two $S_z=-1/2$ states in a row are
necessarily in a spin-1 representation. Thus the braiding does not
commute with the projection: imposing the Jones-Wenzl projector
violates unitarity. Obviously, we can't have this, so the only
alternative is to conclude that $e^{6v}_i$ from Eq.\ \eqref{eSG}
cannot be used to build a braid matrix.

Luckily, this issue is well-understood from a number of points of
view.  When $k$ is an integer, there is another representation of the
Temperley-Lieb algebra which preserves the projection. In the language
of two-dimensional classical statistical mechanical lattice models, this representation is
called the restricted solid-on-solid (RSOS) representation
\cite{ABF,Pasquier}.  In the $S$ matrix language, this representation
describes the scattering of kinks in potential with $k+1$ degenerate
minima, as we discuss in Appendix \ref{app:RSOS}.  In the
quantum-group picture, the RSOS representation is obtained by
``truncating'' the states, so that all are in irreducible and
indecomposable representations of $U_q(sl_2)$.

The presence of the Jones-Wenzl projector means that we do not need to
define the braid matrix on the full tensor product $V^{\otimes N}$,
but only on the restriction/truncation/projection of the space
$V^{\otimes N}$ to the states obeying ${\cal
P}_j^{([k+1]/2)}V^{\otimes N}=0$ for all $j$. This restricted Hilbert
space is our true space of states $V(N)$. The states in $V(N)$ are
conveniently labeled in terms of a series of variables we call ``dual
spins''. (These variables are often called heights; we avoid this here
to avoid confusion with the heights we discuss in the next section.)
The dual spins take on integer values ranging from $1\dots k+1$, and
live {\em between} the strands. Each strand is labeled by the two dual
spins to the left and right of it, which must differ by $\pm 1$. The
key effect of the restriction is the fact that dual spins only range
from $1$ to $k+1$. This is a consequence of not allowing $k+1$
consecutive strands to have spin $(k+1)/2$.

A useful way of understanding the dual spins
comes from treating each strand as being a spin-1/2 representation of
$U_q(sl_2)$, and a dual spin $r$ as being a spin-$(r-1)/2$
representation. Fix the first dual spin $r_1$ to be have
the value $r_1=1$, so it signifies the identity representation of
$U_q(sl_2)$.  Our rules for dual spins mean that a strand next to $r_1$
separates this from a region of $r_2=2$; the region of dual spin 2 can be
next to a region of dual spin $r_3=1$ or $r_3=3$, and so on.  These are
precisely the rules for taking tensor products in $sl_2$: we have
$0\otimes 1/2 = 1/2$, $1/2\otimes 1/2 = 0 \oplus 1$, etc.  In
other words crossing a strand next to the dual spin $r$ is like tensoring
the spin-$(r-1)/2$ representation (the dual spin) with the spin-1/2
representation (the strand). Thus we define the value $(r_i-1)/2$ to be
the total spin of the first $i-1$ strands.  Now imposing the
Jones-Wenzl projector is easy. Forbidding the representation of spin
$(k+1)/2$ is equivalent to forbidding the dual spin $r= k+2$. Note that
our earlier representation $e^{6v}$ can be described in terms of
dual spins as well: the $+$ particle increases the dual spin by $1$ (moving
left to right, say), while the $-$ particle decreases it by $1$. That
the braiding from Eq.\eqref{eSG} can violate the Jones-Wenzl projection
is obvious in the dual spin description: after braiding the value of the
dual spin can reach $k+2$ even if all the initial dual spins are below $k+2$.

To give the explicit RSOS representation of $e_i$ for $k$ integer it
is most convenient to label it by four dual spins $r,s,t,u$, each
ranging between $1$ and $k+1$, with $|r-s|=|s-t|=|t-u|=|r-u|=1$. The
matrix elements of $e_i$ are then \cite{ABF,Pasquier}
\begin{equation}
\begin{picture}(120,40)
\put(-20,13){$e_i=$}
\put(18,0){$r$}
\put(4,14){$s$}
\put(0,0){\line(1,1){40}}
\put(33,14){$u$}
\put(40,0){\line(-1,1){40}}
\put(18,28){$t$}
\put(65,13){$=  \delta_{su}\ $\Large{$\frac{\sqrt{[r]_q [t]_q}}{[u]_q}$}}
\end{picture}
\label{eRSOS}
\end{equation}
where $[h]_q\equiv (q^h - q^{-h})/(q-q^{-1})$. The lines represent the
strands; this picture represents what happens when the $(sr)$ strand
braids with the $(ru)$ strand. After the intersection, the final state
consists of the $(st)$ and $(tu)$ strands.  Since the $S$ matrix (and
the corresponding braid matrix) are non-diagonal, the final state can
be different, namely one can have $r\ne t$ if $s=u$. A very important
thing to note is that if $s$, $r$ and $u$ are between $1$ and $k+1$,
then the matrix elements for $t=0$ and $t=k+2$ vanish because $q^{k+2}=-1$.
integer. Thus if an initial configuration satisfies the Jones-Wenzl
projection, the final one does as well.

Using the matrix of Eq.\eqref{eRSOS} in Eq.\eqref{STL} gives the $S$ matrix of
an integrable 1+1-dimensional field theory; we identify this theory
in the next section. Using the matrix of Eq.\eqref{eRSOS} in Eq.\eqref{Be}
({\it i.e.\/} taking the $\theta\to\infty$ limit of the $S$ matrix) gives the
braid matrix of particles associated with $SU(2)_k$ Chern-Simons
theory.  This is also the braiding of the quasiholes in the
Read-Rezayi states in the fractional quantum Hall effect
\cite{read-rezayi,nayak,slingerland}.  It is important to note that
this representation, Eq.\eqref{eRSOS}, only is useful for $k$ an integer;
otherwise the Jones-Wenzl projection cannot be satisfied (indeed, it
does not exist).

The space $V^{\otimes N}$ for the (wrong) braid matrix (\ref{eSG}) has
dimension $2^N$. Since the actual Hilbert space $V(N)$ is a subspace
of $V^{\otimes N}$, its dimension must be smaller. Finding its size is
a straightforward exercise done in many places; see {\it e.g.\/}  Appendix \ref{app:RSOS}
or Ref.\ \onlinecite{slingerland}. One finds for $N$ large that the
number of states grows as $d^N$. Thus the weighting per loop is indeed
$d$, as the Temperley-Lieb algebra implies it should be.

The results for $SO(3)$ braiding and the BMW braiding can be derived
from the Temperley-Lieb representation, Eq.\eqref{eRSOS}.  There are two
solutions to the Yang-Baxter equation whose $S$ matrices turn into
Eq.\eqref{braidBMW} in the infinite rapidity limit
$\theta\to\infty$. In the classification of Ref.\ \onlinecite{Jimbo},
one solution is associated with the fundamental representation of the
$U_q(A_2^{(2)})$ ($A_2^{(2)}$ is a twisted Kac-Moody algebra), while
the other solution is associated with the spin-$1$ representation of
$U_q(sl_2)$. We will discuss in the next section how both solutions
have been identified as $S$ matrices for two (related) field theories.
The first solution is the most important for us. Written in terms of
the generators $X$ and $E$, it is
\begin{equation}
\widetilde{S}_j = \frac{q^2 e^{\lambda\theta} - q^{-2} e^{-\lambda\theta}  }
{e^{\lambda\theta} - e^{\lambda\theta} } {\cal I}
- \frac{q e^{\lambda\theta} + q^{-1} e^{-\lambda\theta} }
{q^3 e^{\lambda\theta} + q^{-3} e^{-\lambda\theta}  } E_j + X_j
\label{SBMW}
\end{equation}
Taking the $\theta\to\infty$ limit yields the braiding matrix $B^{SO(3)}$ in
Eq.\eqref{braidBMW}.

We can build a representation of the $E_j$ and $X_j$ from the $e_i$ of
a Temperley-Lieb representation by using the relations of 
Eq.\eqref{EXTL}. If the latter obeys the Jones-Wenzl projection, the
resulting $SO(3)$ BMW representation does as well.  In fact, we must
use the representation of Eq.\eqref{eRSOS}, because it turns out that this
yields the only unitary $S$ matrix of the form of Eq.\eqref{SBMW}. The
representation of the states in terms of dual spins therefore applies
to the $SO(3)$ case as well. However, here the rules for adjacent dual
spins are different, since the strand is in the spin-1 representation
of $U_q(sl_2)$. For dual spins $1\dots k-1$, the rules are the same
familiar ones from ordinary $sl_2$ spin-1 representations: {\it e.g.\/} 
$(h-1)/2 \otimes 1 = (h-3)/2 \oplus (h-1)/2 \oplus (h+1)/2$ for $3\le
h \le k-1$.  For dual spins $k$ and $k+1$ we have respectively
$(k-1)/2\otimes 1 = (k-3)/2 \oplus (k-1)/2$ and $k/2\otimes 1 =
(k-2)/2$; in the latter, the representation of spin $k/2$ does not
appear in the right-hand-side.  Note that the states split into two
subsectors, since even dual spins must always be adjacent to even dual
spins, and odd are adjacent to odd.

If one uses the rules given in Appendix \ref{app:RSOS} to count the
number of states for $N$ spin-1 strands, one finds it grows as
$(d^2-1)^N$ at large $N$.  Thus we can indeed interpret $d^2-1$ as the
weight of an isolated loop, as the $SO(3)$ BMW algebra implies.

\section{The 1+1-dimensional field theories}
\label{sec:field}

In order to build our quantum loop gas, we need one more
ingredient. This is to identify the underlying two-dimensional
classical field theory, so that the wave function of the 2+1-dimensional theory 
is given by Eq.\eqref{spsi}. We argued in
Section \ref{sec:qlg} that the 1+1-dimensional version of this
underlying theory will have an $S$ matrix
whose $\theta\to\infty$ limit gives the braid matrix. In this section
we identify the field theories whose $S$ matrices are those in the
last section.

To construct the quantum loop gas directly from the field theory, one
needs to know the action of the two-dimensional classical field
theory. As we will discuss in more detail below, for most of the
theories of interest, the explicit action is fairly difficult to deal
with. However, as discussed in Appendix \ref{app:RSOS}, there are nice Landau-Ginzburg
descriptions. Thus one can define a quantum Hamiltonian in this
language, using the procedure discussed in Refs.
\onlinecite{arovas91} and \onlinecite{ardonne}. 

\subsection{The $SU(2)$ case}

We start with the $SU(2)$ case, arriving at the results of Refs.\
\onlinecite{freedman01,freedman03} and \onlinecite{freedman04} from a slightly
different point of view. The $S$ matrix, Eq.\eqref{STL}, with $e_i$ in the
representation of Eq.\eqref{eRSOS}, describes a field theory which can be
defined in several different ways, which we describe here.

One definition is as the continuum limit of an RSOS
lattice model\cite{ABF}. The degrees of freedom of an RSOS lattice
model are on the sites of a square lattice.  The variables are called
``heights'', and are integers ranging from $1\dots k+2$. Heights on
nearest-neighbor sites must differ by $\pm 1$. The Boltzmann weights
for this model are those of regime III in Ref.\ \onlinecite{ABF}.
This phase is ordered\cite{Huse}. Each ordered state has only two
heights present: one sublattice has all heights $h$ while the other
has all $h +1$. The excitations are the $k$ different kinds of domain
walls between the $k+1$ different ordered states; each wall can be
labeled by the two heights $ h, h \pm 1$ it separates.

It is quite simple to see qualitatively how the $S$ matrix (\ref{STL})
applies to this RSOS height model. The dual spins in the
representation (\ref{STL}) are identified with the ground states of
the height model (there are $k+1$ of each, with the rule that adjacent
ones must differ by $\pm 1$). The strands are identified with the
excitations of the lattice model, the domain walls.  In the absence of
defects the domain walls form non-intersecting loops, just like the
particle worldlines we have described in detail above. The domain
walls are indeed the objects whose scattering is described by the $S$
matrix. In the 1+1-dimensional picture, the excitations can be thought
of as kinks, as discussed in Appendix \ref{app:RSOS}.

Of course there are multiple lattice models with the same continuum
$S$ matrix. The RSOS height model has the advantage that it is
integrable, and that the connection of the $S$ matrix to the lattice
variables is quite intuitive. However, there is another lattice model
in almost the same universality class. We say ``almost'' because some
modifications are required if the space is a torus. This caveat does
not affect the $S$ matrix, and in any case we will not worry about the
torus. The model with the same $S$ matrix is called the lattice $O(n)$
loop model, because at $n$ integer it is $O(n)$ invariant. However, we
are interested in the case $n=d=q+q^{-1} =2\cos(\pi/(k+2))$, so that
$|n|\le 2$. This model can be defined for all $n$ as a gas of self-
and mutually-avoiding loops on the honeycomb lattice with a weight $n$
per loop, in addition to a weight per length of loop.  By writing the
$S$ matrix for the 1+1-dimensional version of the $O(n)$ model in
terms of generators obeying algebraic relations, one in fact can make
at least formal sense of it for all values of $n\le 2$, not just $k$
integer \cite{Zpoly}.  These algebraic relations are equivalent to the
Temperley-Lieb algebra \cite{smirnov92}. The loops are interpreted
heuristically as the world lines of the particles. In these works, no
explicit representation of the Temperley-Lieb algebra is necessary,
but the Jones-Wenzl projection is required to obtain the correct
answer for physical quantities on the cylinder \cite{fendleysaleur}.
Thus when $k$ is an integer one can use the RSOS representation in the
$O(n)$ model as well, although the physical interpretation of this in
the lattice model is not very clear.

(As a side remark, we note that the earlier representation
of Eq.\eqref{eSG} does have a nice heuristic interpretation in the context
of the $O(n)$ lattice model.  One can formulate this model as a model
of oriented loops, where clockwise loops get a weight $q$ and
counterclockwise loops get a weight $q^{-1}$.  Despite the complex
Boltzmann weights, the partition function remains real after summing
over all orientations. The formulation in terms of oriented loops is
useful because this can be mapped onto a model with local
interactions, the six-vertex model with staggered Boltzmann weights
\cite{nienhuis}. The projection mentioned is necessary to get the
correct weighting for loops which wrap around the cylinder. In this
formulation, the $+$ and $-$ particles mentioned at the beginning of
Section \ref{sec:examples} then correspond to the two orientations of
the loop.)

The description in terms of the $O(n)$ lattice model is precisely that
found in Refs.\ \onlinecite{freedman01,freedman03,freedman04}. The
2+1-dimensional lattice model discussed there has a ground-state
wave function of the form of Eq.\eqref{spsi}, where the action ${\cal
S}$ is precisely that of the $O(n)$ loop model with $n=d=q+q^{-1}$. It
was convincingly argued that this model indeed has fractional
statistics, with braid matrix given by Eq.\eqref{Be}.  Thus by our $S$
matrix line of argument, we have arrived at the same conclusion.  This
is therefore strong evidence in favor of our conjecture in Section
\ref{sec:qlg} that the braid matrix of the 2+1-dimensional theory is
related to the $S$ matrix of the corresponding 1+1-dimensional theory.

The theories with the $SU(2)$ RSOS $S$ matrix can be formulated
directly in the continuum, without need for the lattice descriptions
given above. For general $k$, however, there is no simple field-theory
action for these theories, although a heuristic but useful
Landau-Ginzburg description is given in appendix \ref{app:RSOS}. They can also
be defined in terms of constrained fermion
models\cite{constrained-fermion} which realize the Goddard-Kent-Olive
current algebra construction\cite{GKO}. It is difficult to obtain much
information from this formulation, however.
For our purposes, it is most convenient to define
the field theories of interest as perturbations
of a conformal field theory. 
One can define and indeed solve
conformal field theories without a Lagrangian: the Hamiltonian and
states are defined in terms of representations of the Virasoro
algebra. A massive field theory is defined by perturbing the
conformal field theory by a relevant operator.
As shown in Ref.\ \onlinecite{ZRSOS}, 
the $S$ matrix of Eq.\eqref{STL} with $e_i$ given by Eq.\eqref{eRSOS} is
that of a perturbation of the conformal minimal model
with central charge
\begin{equation}
c=1-\frac{6}{p(p+1)}.
\label{cp}
\end{equation}
The desired $S$ matrix describes the perturbation of the conformal
field theory with $p=k+2$ by its least relevant primary field
(known usually as $\Phi_{1,3}$), which has scaling
dimension $2(p-1)/(p+1))$.

Before moving on to the $SO(3)$ case, we wish to note another
complication in the above picture. The first is that the $S$ matrix of
Eq.\eqref{STL} applies to the $O(n)$ model in its dilute phase, where
the energy per unit length is larger than the entropy, so that the
loops cover a small part of the lattice. In order to get a purely
topological field theory, the weight per unit length of loop must be
$1$, so that no length scale is set for the loops\cite{freedman04}.
Such an $O(n)$ model for $n<2$ is in its dense phase,
where entropy wins and the loops cover a set of measure $1$ of the
lattice. However, the braid matrix is not related to the $S$ matrix in
the dense phase, but rather that of the dilute phase.  (The $S$ matrix
in the dense phase has been studied, but due to the non-unitarity of
the model, understanding it precisely is a complicated and somewhat
gruesome story.)  The dense and dilute phases are not dual to each
other; the former has algebraically decaying correlators, while the
latter's decay exponentially.  The same statements can be made in the
context of the height models describing the perturbed minimal models.

The way of understanding this complication is to remember that in the
dilute phase the arguments of Section \ref{sec:qlg} suggest that $S$
matrix really is describing the scattering of the excitations
themselves, {\it i.e.\/} what happens when two world lines braid.
The braiding we are interested in is of the
bare loops, not the renormalized excitations, and this is given by the
$S$ matrix in the dilute phase.
The $S$ matrix in the dense phase is
describing the excitations over the sea of dense loops, which is
important in the 1+1-dimensional case, but not of interest for the
2+1-dimensional braiding. The lesson is
that the braiding should indeed be interpreted as that in the
dilute phase, even though the
topological point is in the dense phase where loops proliferate.

\subsection{The $SO(3)$ theory}

Several field theories with the $S$ matrix of Eq.\eqref{SBMW} were identified and
discussed by Smirnov \cite{smirnov91}. The case we will focus on
here corresponds to a perturbation of a minimal
conformal field theories with central charge of Eq.\eqref{cp}. 
However, for a given $k$, both the minimal model and the perturbation
are different from the $SU(2)$ case. This time, we have
$p=k+1$, and the perturbation is by the $\Phi_{21}$ operator.

It is convenient to use the much-better-known
interpretation of this field theory as the continuum
description of the $Q$-state
Potts model where $Q$ is given by \cite{nienhuis,dotsenko}
$$Q=d^2 = (q+q^{-1})^2 = 4\cos^2\left(\frac{\pi}{k+2}\right).$$ 
The Potts model can defined for all $Q$ in terms of its high-temperature
expansion, where $Q$ becomes a parameter. This definition does not
have local Boltzmann weights for arbitrary $Q$, but for our special
values with $k$ integer, there is a lattice model with the same
high-temperature expansion. This is found by using the original
Temperley-Lieb result of writing the Potts transfer matrix in terms of
generators obeying the algebra of Eq.\eqref{TLalg}, and then using the RSOS
representation, Eq.\eqref{eRSOS}, of these generators. At a particular
coupling where the weights are isotropic, the lattice models are
identical to the RSOS lattice models \cite{ABF,Pasquier} at their
critical point. Thus the Potts critical point is also described by the
conformal field theory with $p=k+1$. Off the critical point, the $S$
matrix of the Potts model with $k$ integer is indeed of the form of Eq.\
\eqref{SBMW}.\cite{smirnov91,pottsnew,CZ,fendleyread} However, when $Q$ is an
integer ($k=2,4,\infty$), this $S$ matrix is diagonal, so the braiding
which follows from it is Abelian. For non-Abelian statistics, we need
to use the Potts model for $Q$ not an integer. We discuss this quantum
loop gas in detail in Section \ref{sec:lattice}.

As opposed to the $O(n)$ model for $n\ne 1$, the Potts model has a
duality relating high to low temperature.  On the lattice, this is a
generalization of the Kramers-Wannier duality of the Ising model
\cite{KW,baxbook}. In the conformal field theory picture, there is a
${\mathbb Z}_2$ symmetry relating the perturbing operator $\Phi_{2,1}$
to $-\Phi_{2,1}$. All the operators appearing in the operator product
expansion of $\Phi_{2,1}$ with itself are irrelevant, so perturbing by
$\Phi_{2,1}$ and $-\Phi_{2,1}$ must be equivalent. The two signs of
perturbation correspond to the low- and high-temperature phases, with
the critical point being the self-dual point.

This duality is a crucial ingredient in interpreting states in the
2+1-dimensional quantum model. We have stressed above how excitations
with non-Abelian statistics can arise in quantum loop gases, where the
ground state is a liquid state, {\it i.e.\/} a superposition of many
states which does not break any symmetries. The key to understanding
how to do this here is to view what we referred to as the
high-``temperature" phase in the classical statistical mechanical
picture as a quantum disordered ground state which we picture to be a
superposition of the excitations of the dual ordered phase, {\it
i.e.\/} the excitations of the classical {\em low}-``temperature"
phase. Recall that in the 2+1-dimensional picture, the weights
measure the amplitude of a particular configuration in a
wave function. This terminology contains somewhat of an abuse of
language: the quantum system is not at high temperature, but
rather the weights of the ground state corresponds to high temperature
in the classical model. We are discussing the
properties of the quantum system only at zero (physical!) temperature.

In the Abelian case, this can be seen quite clearly in Kitaev's model
\cite{kitaev}. Here the underlying classical lattice model is the
Ising model. This is therefore equivalent to both the $SU(2)_1$ model
(based on the $O(1)$ loop gas) and our $SO(3)_2$ model (the $Q=2$-state Potts
model). The loops are simply the domain walls between the Ising
spins, which get a weight $1=(\sqrt{2})^2-1$. The corresponding
2+1-dimensional model is topological when the action ${\cal S}$ in the
wave function of Eq.\eqref{spsi} is of the Ising model at infinite
temperature, where the Ising domain walls have zero energy
per unit length and have proliferated. In the ordered phase, the order operators have
expectation values, and the excitations are created by the disorder
operators\cite{ardonne}. In the disordered phase, the disorder operators get
expectation values, and the excitations are created by the order
operators. The lesson is that when there is a duality, the operator
which creates excitations in one phase is the one which gets the
expectation value in the dual phase\cite{KC,FS}.

To conclude this section, we recall that
there is another model which has an $S$ matrix of the form
of Eq.\eqref{SBMW}. This is the tricritical Potts model, which in conformal
field theory language corresponds to the minimal model with $p=k+2$
perturbed by the $\Phi_{1,2}$ operator. One could presumably build
quantum loop gases based on the tricritical models as well.  Since
in two dimensions, the tricritical point is unstable to perturbations
toward the ordinary critical point, this would presumably hold as well
in the 2+1-dimensional version. Thus such a quantum loop gas would
be near a multicritical point as well.

We also noted above that there is a second $S$ matrix which reduces to
$B^{SO(3)}$ in the $\theta\to\infty$ limit. This $S$ matrix is
associated with a certain perturbation of the $SU(2)_k/U(1)$
``parafermion'' conformal field theories \cite{Fateev}. (The
perturbation is the $\overline{\Psi}_1\Psi_1$ operator, where $\Psi_1$
is the parafermion operators.)  The physics is different for the two
signs of this perturbation. For one sign, one obtains a massive phase,
with this $S$ matrix. For the other sign, one flows to the minimal
model with the central charge given in Eq.\eqref{cp} with
$p=k+1$. This is precisely the critical point of the Potts models!
Moreover, both critical points appear in the same RSOS lattice model
(in the nomenclature of Ref.\ \onlinecite{ABF}, the parafermion
critical point separates regimes I and II, the minimal model separates
regimes III and IV).  Thus our interpretation is that this second $S$
matrix is describing the same quantum loop gas in a region near
another multicritical point.

\section{Lattice models}
\label{sec:lattice}

We can combine all these ingredients to build quantum loop gases on
suitable lattices whose excitations should have non-Abelian statistics. As
discussed in Section \ref{sec:qlg}, the strategy is to build a model
whose ground state is given by a loop gas where the loops have the
correct properties ({\it e.g.\/} a weight of $d$ per loop in the $SU(2)$
case, and a weight of $d^2-1=Q-1$ per isolated loop in the $SO(3)$ case).
Such a lattice model for the $SU(2)$ case was introduced by Freedman, Nayak and Shtengel
\cite{freedman04}. We repeat some of these arguments here, and
then use the $S$ matrix picture to define an analogous model for the
$SO(3)$ case.

\subsection{Criteria for the lattice models}
\label{sec:criteria}

In all of the lattice models discussed here, the Hamiltonian is of the
Rokhsar-Kivelson form, meaning it can be written in terms of a sum of
projection operators:
\begin{equation}
H = \sum_i \lambda_i H_i.
\label{HRK}
\end{equation}
The projection operators $H_i=H_i^2$ are local but not
necessarily commuting. The off-diagonal terms in the $H_i$ must be
ergodic, in the sense that any configuration can be mapped to any
other (with the same values of any globally conserved charges) by
repeated applications of the $H_i$.  To obtain a desired ground state
$|\Psi\rangle$, one must find a set of operators $H_i$ so that
$$H_i|\Psi\rangle =0$$ for all $i$. 
This means that the state $|\Psi\rangle$ is an eigenstate of $H$ with
energy $0$. As long as all the coupling constants $\lambda_i$ are strictly positive,
$\lambda_i>0$, this state $|\Psi\rangle$ is a ground state.
We study models where the solution
of this equation can be written in the form of Eq.\eqref{spsi}: the
basis elements of the Hilbert space can be thought of as a
configuration in a classical two-dimensional lattice model, and the
weight of this configuration can be expressed in terms of a local
action. A key requirement that we will impose is that of locality, {\it i.e.\/} that all the operators $H_i$ act on a finite set of contiguous degrees of freedom.

The degrees of freedom of the models in this section consist of a
quantum two-state variable on each {\em link} of some two-dimensional
lattice. We call these two states occupied and unoccupied. An occupied
link corresponds to the presence of the strand, which one can think of
as being in the spin $1/2$ representation of $U_q(sl_2)$ in the
$SU(2)$ case, and spin 1 in the $SO(3)$ case. An empty site
corresponds to the identity representation. What this means is that at
each vertex, configurations appearing in the ground state must obey
the corresponding fusion rules of
$U_q(sl_2)$. \cite{turaev,reshetikhin,kuperberg,levin03,slingerland}. For example,
three links, in states corresponding to representations $r$, $s$ and
$t$ of $U_q(sl_2)$, touch each vertex of the honeycomb
lattice. Configurations in the ground state must have the identity
representation in the tensor product $r\otimes s\otimes t$. Thus in
the $SU(2)$ case, at each vertex must be touched by zero or two
occupied links. In the $SO(3)$ case, each vertex must be touched by
zero, two, or three occupied links.

\subsubsection{The $SU(2)$ Lattice Loop Models}
\label{sec:lattice-su2}

For the $SU(2)$ case, one needs a set of $H_i$ which annihilates
states with the weighting rules of the $O(n)$ model. In other words, the
ground state must consist of a superposition of configurations where
the strands form self- and mutually-avoiding loops. Moreover, each
loop should have a weight $d$, and to be a purely topological ground state, 
there should be no weight per unit length.  Precisely, the
criteria imposed on the configurations in the ground state at the
purely topological $SU(2)$ point are:\cite{freedman04}
\begin{enumerate}
\item 
The strands form closed non-intersecting loops: {\it i.e.\/} each
vertex has either 0 or 2 links with occupied links touching it.
\item
 If two configurations are related by moving strands around, without
cutting the strands or crossing any other strands, then these two
configurations must have the same weight. In other words, two
topologically identical configurations have the same weight.
\item 
If two configurations are identical except for one having a
closed loop around a single plaquette ({\it e.g.\/} a loop of length 6 on the
honeycomb lattice, length 4 on the square), then the weight of the
configuration without the single-plaquette loop is $d$ times that of
the one with it.
\end{enumerate}
The latter two properties are known as $d$-isotopy\cite{freedman03}.
Note that arbitrarily-sized loops are not directly required to
have weight $d$. Rather, this property follows indirectly by combining the two
latter properties: one can use criterion 2 to
shrink a loop to its minimal size, and then use criterion 3 to
remove it altogether while giving a relative weight $d$ to the
ground-state wave function.

It is now straightforward to find the $H_i$ annihilating a state with
these properties by using locally-defined projection operators. An
explicit expression for the $H_i$ in the $SU(2)$ case on the honeycomb
lattice can be found in Ref.\ \onlinecite{freedman04}. Since we will
not need the explicit Hamiltonian, we will not give it here -- it's
rather ugly, but it does the job.  In these models so far, $d$ is a
parameter that can take on any value, as the Jones-Wenzl projector is not
imposed. To impose this projector on the ground state, one can add an
energy penalty for configurations which violate the projection. This
requires a fine-tuned interaction involving a number of terms
involving $k+1$ spins (or strands) for level $k$.

\subsubsection{The Lattice $SO(3)$ Models}
\label{sec:lattice-so3}

To study the $SO(3)$ theories for arbitrary $k$, we need to work
harder.  The appropriate lattice models are found by imposing criteria
analogous to those of the $SU(2)$ case, but adapted to the spin-$1$
loops. 

\begin{figure}[h!] 
\begin{center} 
\includegraphics[width= .45\textwidth]{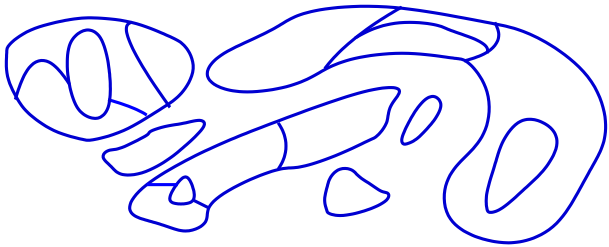} 
\caption{A typical configuration in the spin-$1$ loop model} 
\label{fig:o3loops} 
\end{center} 
\end{figure}

A typical loop configuration in the $SO(3)$ model should look
like that in Fig. \ref{fig:o3loops}.
The lines in this figure represent
``spin-$1$'' particles, so that they correspond to the projected double
lines in the earlier Figs. \ref{fig:proj} and \ref{fig:IXE}.
We thus dub this the spin-$1$ loop model. We
still require that the strands form closed loops. However, as opposed
to the $SU(2)$ case, we must now allow for trivalent vertices, {\it
i.e.\/} the loops are now allowed to branch and merge.  
Thus the spin-$1$ loop model has branching loops, {\it i.e.\/} the configurations 
are nets\cite{levin03}.
In the language of the quantum-group algebra $U_q(sl_2)$, the reason
for the trivalent vertices is that spin $1$ appears in the tensor
product of two spin-$1$ representations.  Equivalently, one can form
an invariant from three spin-$1$ representations.  Pictorially, this
follows from the presence of the $SO(3)$ BMW generator $X$ in Fig.
\ref{fig:IXE}. This generator does not occur in the $SU(2)$ case. In
the quantum-group language, there are three generators here because
the three representations of spin $0,1$ and $2$ appear in the tensor
product of two spin-$1$ representations. Precisely, the projection of
two spin-1 strands onto a spin-zero strand is $E/(Q-1)$, 
and onto a spin-1 strand is $(X-E)/(Q-2)$. 

\begin{figure}[h!] 
\begin{center} 
\includegraphics[width= .2\textwidth]{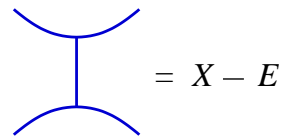} 
\caption{Two trivalent vertices in the $SO(3)$ BMW algebra} 
\label{fig:stdual} 
\end{center} 
\end{figure}

\begin{figure}[h!] 
\begin{center} 
\includegraphics[width= .3\textwidth]{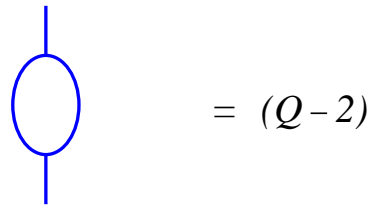} 
\caption{Removing one loop in the spin-1 loop model} 
\label{fig:q-1loop} 
\end{center} 
\end{figure}

In the spin-$1/2$ loop model for the lattice $O(n)$ model
\cite{freedman04}, the Temperley-Lieb relation $e_i^2=d e_i$ implies
that each loop receives a weight $d$.  We can understand the somewhat
more intricate analogous properties of the spin-$1$ loop model by
using the $SO(3)$ BMW algebra. The relation $E_j^2=(Q-1)E_j$
implies that isolated loops in the spin-$1$ model receive a weight of
$Q-1=d^2-1$. Because trivalent vertices occur here, however, all loops
need not be isolated. The projector onto spin-$1$ is proportional to
$X-E$, so we associate this with two neighboring trivalent vertices,
as indicated in Fig.\ \ref{fig:stdual}.
Several properties of the loop gas follow from this.
The relation $(X_j-E_j)^2 = (Q-2)(X_j-E_j)$ means that a configuration
with a loop with just two lines emanating from it has a weight $Q-2$
times the configuration with the loop removed. This is illustrated in
Fig. \ref{fig:q-1loop}.
Moreover, because $(X_j-E_j)E_j=0$, no graph can contain any loop with
just one external line attached to it. We call such a forbidden loop a
``tadpole''.

We must work harder to find the weight of more complicated
configurations in the loop model.  To make the answer precise, we use
a two-dimensional classical lattice model which has a loop expansion
with the desired properties. As we discussed in Section
\ref{sec:field}, for the spin-$1$ loop model this should be the
$Q$-state Potts model, since its $S$ matrix gives the desired braid
matrix.  The desired loop expansion is the {\em low}-temperature
expansion of the Potts model. 

Let us first describe the low-temperature loop expansion 
for $Q$ integer, where the Potts models are
defined by placing a ``spin'' $\sigma_i$ taking values $1\dots Q$ at
the sites $i$ of a lattice. As it is well known, the interaction for a Potts model
depends only on whether nearest-neighbor spins are the same or
different, so that the Boltzmann weight for a link with spins $\sigma_i$
and $\sigma_j$ at its ends is
$$e^{K(\delta_{\sigma_i \sigma_j}-1)}.$$ 
The low-temperature expansion is given by first expressing each
configuration of Potts spins in terms of its domain walls. The domain
walls reside on the links of the dual lattice, and separate regions on the direct lattice of spins of
different values. Each link crossing a domain wall has weight
$e^{-K}$, while each link without a domain wall has weight $1$.
The Boltzmann weight of a given spin
configuration depends only on the length $L$ of its domain walls:
$L$ is the number of links on the dual lattice with walls on them.
The Boltzmann weight of a configuration is then $e^{-KL}$. A weight of $1$
per unit length corresponds to infinite temperature in this classical
lattice model.

\begin{figure}[h!] 
\begin{center} 
\includegraphics[width= .1\textwidth]{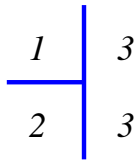} 
\caption{A trivalent vertex in the 3-state Potts model} 
\label{fig:trivalent} 
\end{center} 
\end{figure}

By definition, the domain walls must form loops surrounding groups of like Potts
spins. These loops can intersect, but no tadpoles can occur.
For example, the trivalent
vertex given in Fig. \ref{fig:trivalent} occurs for $Q>2$.

Different
configurations of spins can have the same domain-wall configuration:
{\it e.g.\/} there are $Q$ configurations with no domain walls, and $Q(Q-1)$
configurations with a loop of length $4$ surrounding a given site.  In
general, the number of spin configurations which have the same loop
configuration ${\cal L}$ is called the number of $Q$-colorings
$\chi_Q^{}({\cal L})$ (see {\it e.g.\/} Ref.\ \onlinecite{FK}).  Imagine each
region of like spins to be shaded some color. The number $\chi_Q^{}$ is then
the number of ways this shading can be done with $Q$ colors so that no
two adjacent regions have the same color (regions which meet only at a
point are not considered to be adjacent).  

The partition function of the Potts model can therefore be written as
\begin{equation}
Z= \sum_{\cal L} e^{-KL} \chi_Q^{}({\cal L})
\label{low}
\end{equation}
where the sum is over all {\em distinct} loop configurations: the
multiple spin configurations with the same loop configuration is
accounted for by the factor $\chi_Q^{}({\cal L})$. A typical loop
configuration ${\cal L}$ looks like that in fig.\ \ref{fig:o3loops}.
Because $\chi_Q^{}({\cal L})$ vanishes for any configuration with a tadpole,
or a strands with dangling end, we need not include such
configurations in the sum. This expansion is a
useful description of the ferromagnetic ($K>0$) Potts model at low
temperature.  It is important to note that this is not the only loop
expansion of the Potts model: another expansion is in terms of the
(self- and mutually-avoiding) loops surrounding the clusters in the
high-temperature expansion. \cite{FK,nienhuis,saleur}.

\begin{figure}[h!] 
\begin{center} 
\includegraphics[width= .4\textwidth]{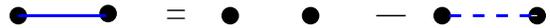} 
\caption{The recursion relation for the chromatic polynomial} 
\label{fig:recursion} 
\end{center} 
\end{figure}

The low-temperature expansion of the partition function of the Potts model, Eq.\eqref{low}, applies
to any $Q$ when $\chi_Q^{}({\cal L})$ is the {\em chromatic polynomial} of
the graph dual to ${\cal L}$.\cite{pottsnew} The graph dual to ${\cal
L}$ is defined with a node corresponding to each loop, and a line
between two nodes when the corresponding loops share a boundary.
In terms of the Potts spins, each node in this graph corresponds to
a region of like spins, and a line between two nodes means that
corresponding regions are adjacent. The chromatic polynomial reduces
to the number of colorings of the graph when $Q$ is an integer, but
can be defined for all $Q$ by a recursion relation. Consider two
nodes connected by a line $l$ ({\it i.e.\/} two loops sharing a boundary
in the original picture). Then define ${\cal D}_l{\cal L}$ to be the
graph with the line deleted, and ${\cal C}_l{\cal L}$ to be the graph
with the two nodes joined into one. Then we have
\begin{equation}
\chi_Q^{}({\cal L}) = \chi_Q^{}({\cal D}_l{\cal L}) - \chi_Q^{}({\cal C}_l{\cal L}).
\label{recursion}
\end{equation}
We represent this pictorially in fig.\ \ref{fig:recursion},
where a node represents each loop, a solid line between two nodes indicates
that the corresponding two loops are adjacent, and a dashed line
indicates that two formerly independent loops are now merged ({\it i.e.\/}
the occupied links separating them are removed). 
This is fairly obvious in the coloring description: $\chi_Q^{}({\cal D}_l
{\cal L})$ includes all graphs in $\chi_Q^{}({\cal L})$, but also has graphs
where the two nodes connected by line $l$ have the same
color. There are $\chi_Q^{}({\cal C}_l{\cal L})$ of the latter so we need to
subtract these off to get the recursion relation. For any loop
configuration ${\cal L}$, one can apply (\ref{recursion}) repeatedly
until one reaches graphs with all isolated nodes. A graph with ${\cal
N}$ isolated nodes has $\chi_Q^{}=Q^{\cal N}$. We will give explicit
examples of how this
works in 
Subsection \ref{subsec:chrom}.

The criteria for the Potts loop model to describe the ground state of the
$SO(3)$ loop gas on the lattice are therefore
\begin{description}
\item{$1^\prime$.} The strands form closed loops, but now we allow trivalent
vertices. 
\item{$2^\prime$.} If two configurations are related by moving strands around without
cutting the strands or crossing any other strands, then these two
configurations must have the same weight. In other words, two
topologically identical configurations have the same weight.
\item{$3^\prime$.} Each loop configuration ${\cal L}$ receives a
weight $\chi_Q^{}({\cal L})$. For example, if two configurations are identical
except for one having a closed loop around a single plaquette (a loop
of length 6 on the honeycomb lattice, length 4 on the square), then
the weight of the configuration without the single-plaquette loop is
$Q-1$ times that of the one with it.
\end{description}
Criterion $2^\prime$ is the same as criterion $2$ in the
$SU(2)$ case; this is the requirement of topological invariance.
Criterion $1^\prime$ is the generalization of criterion $1$, allowing
for trivalent vertices in the $SO(3)$ case. Criteria $3^\prime$ is
the appropriate generalization of criterion
$3$. However, implementing this using a local Hamiltonian requires a
little work, which we will now describe.

\subsection{A Hamiltonian yielding $SO(3)$ statistics}
\label{subsec:chrom}

In the previous Subsection we set out the criteria which the
ground-state wave function for the $SU(2)$ and $SO(3)$ models must
obey. Here we describe a Hamiltonian of the form of Eq.\eqref{HRK} for the
$SO(3)$ case, which has a ground state with the weights of the
(low-temperature) Potts loop gas. This is tantamount to finding a set
of projection operators $H_i$ which annihilate the desired ground
state, and which result in an ergodic Hamiltonian.

There are two types of $H_i$ operators in our Hamiltonian. The
simplest type have purely potential terms, diagonal terms where
$H_i=1$ on some basis elements of the Hilbert space, and zero on the
remaining elements. Such terms thus allow us to satisfy criterion
$1^\prime$: we give a positive potential to any vertex which has only
one occupied link touching it. Since the ground state has energy zero
by construction, any state on which these $H_i$ are non-zero cannot be
part of the ground state, as long as there are no off-diagonal terms
which mix this state with an allowed one.

The second type of term contains off-diagonal elements, and are needed to
ensure that different basis elements have the desired relative
weighting in the ground state. A state of the form $|{\cal L}\rangle +
\alpha|{\cal L}'\rangle$ is annihilated by 
\begin{equation}
H_i=\begin{pmatrix} \alpha&-1\\-1& \alpha^{-1}\end{pmatrix}
\label{projform}
\end{equation} so that
{\it e.g.\/} $H_i|{\cal L}\rangle= \alpha|{\cal L}\rangle - |{\cal
L}'\rangle$.  Since we know each state ${\cal L}$ in the ground state
receives a weight $\chi_Q^{}({\cal L})$, we need $\alpha=\chi_Q^{}({\cal
L}')/\chi_Q^{}({\cal L})$.  

However, if we include an element $H_i$ in the Hamiltonian for any
pair of states ${\cal L},{\cal L}'$, the Hamiltonian will be non-local
for several reasons. An obvious one is that if we include an $H_i$ for
any pair of configurations, the off-diagonal terms are clearly
non-local. We can easily solve this problem by setting $\lambda_i=0$
for any $H_i$ involving an ${\cal L}$ and a ${\cal L}'$ whose
differences are non-local. In other words, we only allow off-diagonal
terms in $H$ which map a given ${\cal L}$ to an ${\cal L}'$ which
differs from ${\cal L}$ only in some small neighborhood (say the links
on a given plaquette). 
While this is a necessary condition for a local
Hamiltonian, it is not sufficient for this model. The reason is that
evaluating $\chi_Q^{}({\cal L})$ for a given ${\cal L}$ is a non-local
operation: it requires knowing the entire cluster of occupied links.
However, even though the overall $\chi_Q^{}({\cal L})$ needs to be
determined globally, the ratio $\chi_Q^{}({\cal L})/\chi_Q^{}({\cal L}')$ in some
cases depends only on local difference between ${\cal L}$ and ${\cal
L}'$, not their global form. Since the Hamiltonian only depends on
this ratio, we can find a local Hamiltonian if we can find
such pairs ${\cal L}$ and ${\cal L}'$.

Let us first describe how to implement criterion $2^\prime$, which says
that if two configurations ${\cal L}$ and ${\cal L}'$ are
topologically identical, then they have the same weight in the ground
state. By definition, if ${\cal L}$ is topologically identical to
${\cal L}'$, we have $\chi_Q^{}({\cal L})=\chi_Q^{}({\cal L}')$. Thus including
$H_i$ of the form of Eq.\eqref{projform}, with $\alpha=1$ will insure the
proper weighting. These terms will be local if we require ${\cal L}$
and ${\cal L'}$ to not only be topologically identical, but completely
identical except on the links around one plaquette.

\begin{figure}[h!] 
\begin{center} 
\includegraphics[width= .28\textwidth]{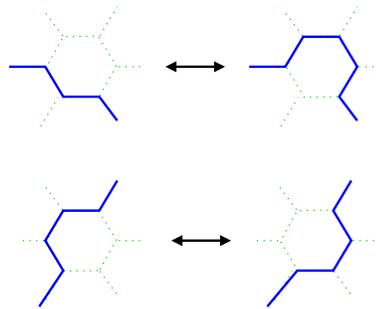} 
\caption{Plaquettes with two external occupied links} 
\label{fig:isotopy2} 
\end{center} 
\end{figure}

To make these terms in the Hamiltonian more specific, 
let us work henceforth  on the honeycomb lattice, 
so that we do not have to worry about 
loops which touch at only a point. Consider a single plaquette,
where some but not all of its six links are occupied. The simplest
possibility allowed by criterion $1^\prime$ is then for two of the
six outside links touching the plaquette to be occupied, as in all the
configurations in Fig. \ref{fig:isotopy2}.
For each configuration of the two outside occupied links (15
possibilities in all), there are two topologically-identical
configurations on each plaquette.  We thus include in $H$ the
$\alpha=1$ projectors which include flips between the two
topologically-identical configurations; two of the flips are
illustrated in Fig.\ \ref{fig:isotopy2}.  These $H_i$ are local,
involving only states on 12 links: the six on the plaquette and the
six touching it. 

\begin{figure}[h!] 
\begin{center} 
\includegraphics[width= .45\textwidth]{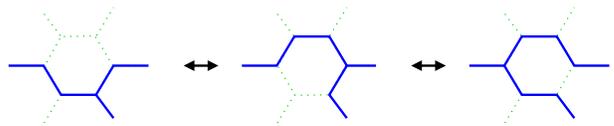} 
\caption{Plaquettes with three external occupied links} 
\label{fig:isotopy3} 
\end{center} 
\end{figure}

\begin{figure}[h!] 
\begin{center} 
\includegraphics[width= .45\textwidth]{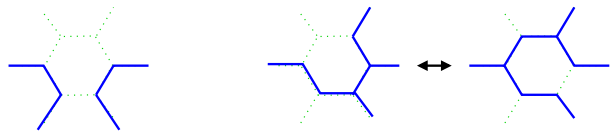} 
\caption{Plaquettes with four external occupied links} 
\label{fig:isotopy4} 
\end{center} 
\end{figure}

This idea can readily be generalized to plaquettes
with more of the outside links occupied. If there are three outside
links occupied, then there are three topologically-identical
configurations on each plaquette, as illustrated in one case in Fig.\
\ref{fig:isotopy3}. 
We thus include $\alpha=1$ projectors which flip
between any pair of topologically-identical configurations. 
For four outside
lines, there are two possibilities. The first type of configuration 
shown
in Fig.\ \ref{fig:isotopy4} has no topologically-identical partner,
while the second type has one.
\noindent
We thus include $\alpha=1$ projectors
for all configurations of the latter type. For five or six outside
lines, we include no projectors. 
By repeatedly applying the $H_i$ described in Figs.\ \ref{fig:isotopy2},
\ref{fig:isotopy3} and \ref{fig:isotopy4}, we implement criterion
$2^\prime$.  

As if this Hamiltonian weren't already complicated
enough, we now need to implement criterion $3^\prime$.
The $H_i$ described in Figs.\ \ref{fig:isotopy2} all map between
topologically-identical configurations. To map between 
topologically-distinct configurations, we need still more $H_i$ operators.
These are still of the form of Eq.\eqref{projform}, but to ensure
different configurations have the correct relative
weight, the $\alpha$ are not necessarily equal to $1$.
Again, we focus on a single plaquette, but here we consider only
plaquettes with all six links occupied. The $H_i$ depend of course on
which of the outside links are occupied. The cases with $0$, $1$, $2$,
and $3$ occupied outside links and all internal links occupied
are easy to implement. We have:
\begin{description}
\item{0.} When there are
no outside links occupied, then a plaquette with all six links
occupied forms an isolated loop. If we remove this loop, the resulting
configuration has weight $Q-1$ relative to the configuration with the
loop. This can be implemented with an $H_i$ with $\alpha=Q-1$,
where ${\cal L}$ is the configuration with the isolated loop and ${\cal
L}'$ is the configuration without it. This is illustrated in fig.\ \ref{fig:isolatedloop}.
\begin{figure}[h!] 
\begin{center} 
\includegraphics[width= .35\textwidth]{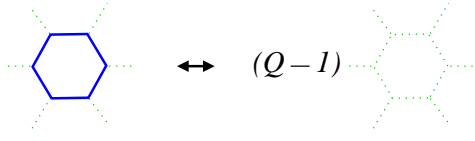} 
\caption{Removing or adding an isolated loop} 
\label{fig:isolatedloop} 
\end{center} 
\end{figure}
\item{1.} If there is just one occupied link, this is a tadpole, and
is forbidden in the Potts loop expansion and hence the ground
state. Thus we add a potential-only term ({\it i.e.\/} $H_i =1$ on a
tadpole). Tadpoles comprised of loops larger than a plaquette end up
being forbidden by using the isotopy: applying the Hamiltonian enough
will shrink a given loop to a single plaquette, and potential here
will then exclude it from the ground state.
\item{2.} If there are two occupied outside links connected to loop,
the configuration with the loop removed has relative weight $Q-2$, as
illustrated earlier in Fig.\ \ref{fig:q-1loop}. This is implemented by
$H_i$ with $\alpha=Q-2$, where ${\cal L}$ is the configuration with
the loop, and ${\cal L}'$ is a configuration with the same external
lines but without the complete loop. As illustrated in Fig.\
\ref{fig:isotopy2}, for a given pair of occupied outside links, there
are two allowed configurations on the plaquette with no loop. We can
include a $H_i(\alpha=Q-2)$ for either or both of these two allowed
configurations.
\item{3.}  If there are three occupied outside links connected to loop,
the configuration with the loop removed has relative weight
$Q-3$. We therefore use an $H_i$ with $\alpha=Q-3$. Here ${\cal L}'$
can be any one of the three configurations illustrated in Fig.\ 
\ref{fig:isotopy3}. 
\end{description}

A Hamiltonian comprised of the $H_i$ we have constructed so far has
ground states with the correct relative weightings. However, it is not
ergodic: there are multiple distinct ground states which are not
related by any of the above off-diagonal terms. For example, a
configuration with four adjacent plaquettes with all six links
occupied, as illustrated in Fig.\
\ref{fig:graph4} below, is annihilated by all the $H_i$ we have discussed
so far. Thus we need more terms in $H$ so that this state alone is not
a ground state. Removing one of these loops (all of which have at
least four occupied outside links) is not as simple as with three or
less occupied links. We need to use the recursion relation
Eq.\eqref{recursion} for the chromatic polynomial to find $H_i$ which
remove these loops.

To find these terms, it is convenient to use the graphical
representation of a loop configuration, as defined above. Knowing the
graph of ${\cal L}$ is sufficient to find its chromatic polynomial
$\chi_Q^{}({\cal L})$.
In terms of these graphs, the recursion relation Eq.\eqref{recursion} can
be represented as in Fig.\ \ref{fig:recursion} above. We can use this
relation to easily rederive the $H_i$ acting on plaquettes with $0$,
$2$ and $3$ occupied outside links. For example, the graph for an
isolated node is precisely that on the left-hand-side of Fig.\
\ref{fig:recursion}. Applying Eq.\eqref{recursion} once, and then using
the fact that an isolated node gives a factor $Q$ to the chromatic
polynomial, gives the desired relative weighting $Q-1$. The equalities
are meant as between the corresponding chromatic polynomials.

Finding $H_i$ which remove a loop with four external lines is
trickier. One can apply the recursion relation, but one can get graphs
which do not correspond to any configuration on the honeycomb lattice
given by changing links on the plaquette from occupied to
unoccupied. Consider the first graphical representation of four
adjacent hexagons in
Fig.\ \ref{fig:graph4}. 
\begin{figure}[h!] 
\begin{center} 
\includegraphics[width= .48\textwidth]{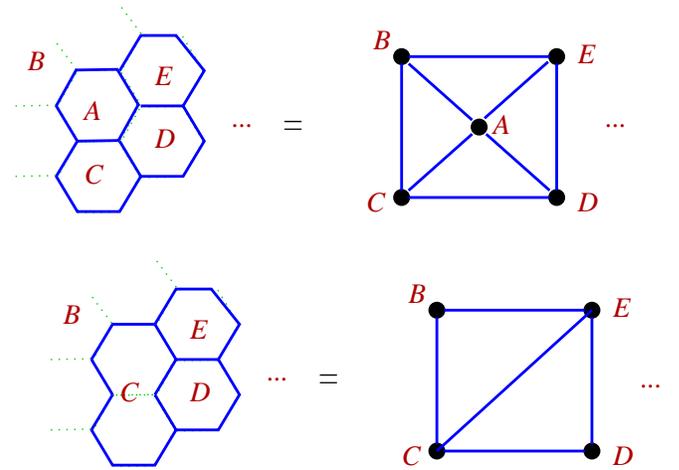} 
\caption{Graphical representations of two configurations with five loops} 
\label{fig:graph4} 
\end{center} 
\end{figure}
In this graph, we allow the nodes $B$, $C$, $D$ and $E$ to be attached
to other nodes, but node $A$ touches only the four in the picture. We
apply the recursion relation once to remove one of the lines attached
to node $A$. This gives a perfectly valid relation for the
corresponding chromatic polynomials, but there is no loop
configuration corresponding to the graph with one line removed.  For
example, say we remove the line from $A$ to $D$: no loop
configuration corresponding to this graph can be drawn on the
honeycomb lattice.

\begin{figure}[h!] 
\begin{center} 
\includegraphics[width= .45\textwidth]{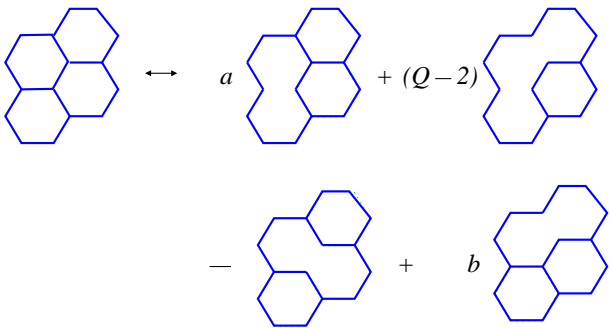} 
\caption{Removing one loop of a four-hexagon cluster} 
\label{fig:fourhex} 
\end{center} 
\end{figure}

To define a Hamiltonian, we need a relation between valid loop
configurations, not just different chromatic polynomials. We can
relate the two loop configurations in Fig.\ \ref{fig:graph4}. In the
first graph, we use the recursion relation to remove the line from $A$
to $D$ and then the line from $A$ to $E$. Now node $A$ has
lines only to nodes $B$ and $C$, and corresponds to the situation of
Fig.\ \ref{fig:q-1loop}. We can now remove node $A$
altogether, multiplying the result (the square involving $B,C,D$ and
$E$) by $Q-2$. This square defines a chromatic polynomial, but there
is no corresponding loop configuration. We can however, relate it to
the second configuration in Fig.\ \ref{fig:graph4}; the same
square graph arises from using the recursion relation to remove the
link from $C$ to $E$. Combining the two, we obtain
the relation in Fig.\ \ref{fig:fourhex} with $a=Q-2$ and $b=-1$.
Note that the first and last configurations are topologically
identical, so we can choose any value of $a$ if we set $b=Q-3-a$. Thus
a more symmetric Hamiltonian will result if we choose
$a=b=(Q-3)/2$. As a check on this relation, we can look at the special
case where none of the nodes $A,C,D,E$ are connected to any others
(i.e.\ the four hexagons $A,C,D,E$ are surrounded by region $B$).
One can then easily verify that both sides are equal to
$(Q-1)(Q-2)(Q-3)^2$.

We can then include an $H_i$ with ${\cal L}'$ the sum of loop
configurations on the right-hand-side of Fig. \ref{fig:fourhex}.  One
can define analogous relations to define new ${\cal H}_i$ which
further reduce the number of loops on the right-hand-side of Fig.\
\ref{fig:fourhex}. Although we have not proven so, we believe
that proceeding in this fashion one can define local $H_i$ which
result in an ergodic Hamiltonian. This should also be possible on
other lattices, but presumably be even more complicated, since one
must worry about loops which touch at only a point.  Obviously, there
is no conceivable way such finely-tuned Hamiltonians could be realized
in nature. However, the fact that they are local makes it at least
possible that there exists a more natural Hamiltonian in the same
universality class.

\section{The phase transition}
\label{sec:phase}

Since we have identified the classical field theories which
describe the ground state of the quantum loop gas, we can determine
the phase structure of the latter. There is a major subtlety in doing
so. So far, we have been discussing properties of the
wave function $\psi$. However, in determining equal-time correlation functions
of the quantum system, the functional integral is weighted by $|\psi|^2$.
In terms of the action of the corresponding classical model, we have from
Eq.\eqref{spsi},
$$|\psi(s)|^2 = e^{-S(s)-S^*(s)} $$ for a configuration $s$ in the
classical model.  Thus when we are computing for example correlators
in the quantum model, we need to {\em square} the Boltzmann weights of
the classical model. In a model where the loops themselves are the
degrees of freedom (such as those discussed in Section
\ref{sec:lattice}, the ones without the Jones-Wenzl projector
imposed), this means the weight per loop must be squared.  In
particular, this means loops in the $SU(2)$ model get weight $d^2$,
while isolated loops in the $SO(3)$ model have weight $(d^2-1)^2$.

The phase structure depends on the weight per length of the loops. 
Let us first review what happens in a quantum eight-vertex model
\cite{kitaev,ardonne}. There the degrees of freedom in the ground
state are arrows on the links of the square lattice obeying the
eight-vertex condition (an even number of arrows pointing in at each
vertex). The loops are given by following (say the up and
right-pointing) arrows around the lattice; the eight-vertex condition
ensures the loops are closed.  One can rewrite these degrees of
freedom as Ising spins at the center of each plaquette, and the loops
then form domain walls around the spins.  The purely topological
(Kitaev) point\cite{kitaev}, corresponding to our $SU(2)_1$ and
$O(3)_2$ models, has equal amplitude for all
configurations. This is indeed infinite temperature for the Ising
spins. By including a weight per length of the loop, one can move away
from the Kitaev point. As detailed in Ref.\ \onlinecite{ardonne}, there is
eventually a phase transition to an ordered phase.

The interesting question to answer is if this is a critical point or a
first-order phase transition. In the quantum eight-vertex model of
Ref.~\onlinecite{ardonne}, the transition is second order.  In the
analogous transition in the $SU(2)$ case \cite{freedman01,freedman03},
each loop will get a weight $d^2$, so we can view this as a loop model
with $n_{\rm eff}=d^2$. The $O(n_{\rm eff})$ model has a critical
point only when $n_{\rm eff}=4\cos^2(\pi/(k+2))\le 2$, so this occurs
only for $k=1$ and $k=2$. Only the latter is non-Abelian.

The analogous result in the classical Potts model is that the phase
transition at the self-dual point is second-order if $Q\le 4$,
first-order for $Q>4$. This result, however, cannot instantly be
applied to the 2+1-dimensional case, because of the weighting by
$|\psi|^2$: the phase structure is that of the classical loop gas
where each configuration is weighted by $(\chi_Q^{}({\cal
L}))^2$. This loop gas does not seem to have been studied before, so
we do not know the answer.  However, we can make a simple
conjecture. The configurations of this squared loop gas are of course
the same as those of the Potts loop gas; only the weight per loop has
been squared. One might therefore hope that the phase transition of
the squared loop gas is in the same universality class as the Potts
loop gas at some $Q_{\rm eff}$. The simplest possibility is to assume 
that $Q_{\rm eff}$ gives isolated loops in the squared loop gas
the correct weight $(Q-1)^2$. This amounts to a $Q_{\rm eff}$ of
\begin{equation}
Q_{\rm eff} - 1 = (Q-1)^2 =(d^2-1)^2 .
\end{equation}
The weight per unit length is also changed, but this only changes the
location of the critical point, not its type. Thus we conjecture that
there will be a critical point in the quantum loop gas if $Q_{\rm
eff}\le 4$, and a first-order transition otherwise.

If this conjecture is true, critical points occur for $k=2,3$ in the
$SO(3)$ model (as noted above, $k=1$ is a trivial theory here). The
$k=3$ model has non-Abelian statistics. As noted at the end of section
\ref{subsec:jones}, this ``Lee-Yang'' model is the simplest model of
non-Abelian statistics, since it has only one kind of strand.  We have
thus shown that in the quantum loop gas which realizes particles with
these statistics, there exists a critical point separating the
topological phase from an ordered one.

However, there's a catch. The Jones-Wenzl projector needs to be
imposed separately to the loop models discussed in Section
\ref{sec:lattice}, at least when space is an annulus or a torus.
This presumably amounts to a relevant operator at the critical
point \cite{freedman04}. Thus in the non-Abelian case even at $k=3$, 
reaching the critical point from the topological phase requires tuning
another parameter away.

This is in harmony with some of our early observations.  At the end of
Section \ref{subsec:bmw} we noted that for the $SO(3)$ model with
$k=3$, one can use the Jones-Wenzl projector to remove $X$ vertex,
leaving only self-avoiding loops. The $SO(3)_3$ model with the
projection is therefore equivalent (at least locally) to the $SU(2)_3$
model.  The latter does {\em not} have a critical point, even without
the Jones-Wenzl projector, because $k=3$ in $SU(2)$ case corresponds
to an $O(n_{\rm eff})$ loop model with $n_{\rm eff}>2$. This is a
strong indication that imposing the Jones-Wenzl
results in a relevant perturbation of the critical theory.


\begin{acknowledgments} 
 
We are grateful to E.~Ardonne, M.~Freedman, S.~Kivelson, C.~Nayak,
K.~Shtengel, M.~Stone and X.G.~Wen for many illuminating conversations
on this and related work. We also thank C.~Nayak for
many useful comments on this manuscript. This work was supported in part by
the National Science Foundation through the grants NSF-DMR-0442537 at
the University of Illinois, NSF-DMR-0412956 at the University of
Virginia, and NSF-PHY-9907949 at the Kavli Institute for
Theoretical Physics, UCSB, where we were participants at the
program on {\em Exotic Order and Criticality in Quantum Matter}. We
thank the organizers, the staff, and the director for their
kind hospitality at the KITP. We likewise thank the American Institute of
Mathematics and the organizers of the inspiring conference on
topological order there. The work of P.F.\ was also supported by
the DOE under grant DEFG02-97ER41027.
 
\end{acknowledgments} 

\appendix
\section{The Landau-Ginzburg description of the 1+1-dimensional theories}
\label{app:RSOS}

In this appendix we explain how the field theories discussed in
section \ref{sec:field}, and their $S$ matrices discussed in section
\ref{sec:examples}, have a nice description in terms of a
Landau-Ginzburg effective field theory.

As shown in Ref.\ \onlinecite{ZamoLG}, a simple Landau-Ginzburg
description of the minimal model of conformal field theory with
central charge of Eq.\eqref{cp} is in terms of a single scalar field
$\phi$ with potential $\phi^{2(p-1)}$. The critical point of the Ising
model is the $p=3$ case, and the tricritical point is the $p=4$ case,
and the corresponding $\phi^4$ and $\phi^6$ potentials have long been
known. The critical point of the $SU(2)_k$ case (the continuum limit
of the $O(n)$ or restricted height models) has $p=k+2$. The critical
point of the $SO(3)_k$ case (the continuum limit of the Potts or
dilute $A_{k+1}$ models) has $p=k+1$.

The primary fields of the conformal field theory also can be described
in terms of $\phi$, so the massive field theories of
interest have a Landau-Ginzburg description as well. For the $SU(2)_k$
case, the effective description of the $\Phi_{1,3}$ operator is
$\phi^{2(p-2)}$. For the Ising model, this is indeed the usual
$\phi^2$ mass term.  As is well known, this field theory has a single
quasiparticle with $S=-1$.
In this case, $e_i$ must act on a one-dimensional space,
and it is just a number; for $p=3$, $e_i=1$.  
This also follows from imposing the Jones-Wenzl projector:
for $k=1$, one imposes ${\cal P}^{(1)}_i(d=1)=I-e_i =0$, which indeed
gives $e_i=I$.  Thus the statistics for $SU(2)_1$ is Abelian
\cite{freedman01,freedman03}.

For higher values of $p$, perturbing a $\phi^{2(p-1)}$ potential by
$\phi^{2(p-2)}$ seems to induce a flow to the minimal model with $p$
decreased by $1$. This flow is indeed known to occur for perturbations
of one sign of $\Phi_{1,3}$, see Ref. \onlinecite{ludwigcardy}. In the $O(n)$
language, this perturbation corresponds to flowing into the dense
phase. In the models of Ref.\ \onlinecite{ABF} this is called regime
IV. This is not what we want; this is a massless field theory with the
problematic $S$ matrix. Instead, we want to perturb by the
same operator $\Phi_{1,3}$ but with the opposite sign; the
perturbations are not the same because there is no symmetry in the
conformal field theory (except in the Ising case) which sends
$\Phi_{1,3}$ to $-\Phi_{1,3}$.  With this sign of the perturbation,
the potential must renormalize to include extra terms. Since we know
from the exact results\cite{ABF,Huse} that the
lattice models have $p-1$ ground states, we must fine-tune the
potential to achieve this.  This is very familiar from the tricritical
Ising model $p=4$. One sign of the perturbation moves the system along
its first-order transition line, while the other sign causes a flow to
the ordinary Ising critical point. The Landau-Ginzburg potential along
the first-order line indeed consists of a $\phi^6$ potential tuned to
have three degenerate minima.

Let us focus on $p=4$ case in more detail.  The Landau-Ginzburg
potential for the tricritical Ising model along its first-order
transition line along is $\phi^6 + a \phi^4 + b\phi^2$, with $b=-6a^2$
so that there are three degenerate minima at
$\phi=0,\pm\sqrt{2a}$. With such a potential, the low-energy
configurations in the two-dimensional classical model consist of
regions of these three vacua.  The loops are domain walls between
different vacua. When there are restrictions on which vacua can be
adjacent to each other, the allowed domain walls are restricted as
well. What happens here is that the vacuum $+\sqrt{2a}$ is not allowed
to be next to the vacuum $-\sqrt{2a}$. Hence, there must be a region
of vacuum $0$ in between. In the 1+1-dimensional description in
terms of quasiparticles, the particles are kinks interpolating between
adjacent vacua, and the domain walls are their world lines. 

One can now count the ``number'' of particles. Say the left end of the
system is in the vacuum $0$. Then the space of states for one particle
$V(1)$ is two-dimensional; it consists of a kink going from $0$ to
$+\sqrt{2a}$, and one going from $0$ to $-\sqrt{2a}$. However, the
space $V(2)$ is also two-dimensional: the reason is this
restriction that the two vacua $\pm \sqrt{2a}$ cannot be next to each
other.  Thus, if we have a kink going from $0$ to $+\sqrt{2a}$, the
next kink can only go back to $0$ again.  The dimension of the allowed
Hilbert space for $N$ ``particles", $V(N)$ (with the boundary
condition of vacuum 0 at one end), is therefore $2^{[(N+1)/2]}$, where
$[x]$ is the integer part of $x$.This reduction of the Hilbert space, with respect to the
standard free particle Fock space, is the hallmark of non-Abelian
statistics.  Using the standard definition of the ``number of
particles" as ln(dim$(V(N)))/N$, here we find $\sqrt{2}$
This is indeed the correct value of $d$ for $k=2$, and gives a precise
meaning to the statement that $d$, the weight per loop, can be
interpreted as counting the ``number of particles going around a given
loop" (albeit this ``number" is $\sqrt{2}$ !).

Thus we see how these restricted kinks in 1+1 dimensions go
hand-in-hand with non-Abelian statistics in 2+1 dimensions. In the
Moore-Read theory of the $\nu=5/2$ fractional quantum Hall effect
\cite{moore92}, the quasiparticles have non-Abelian statistics, and
the number of states obeys the same formula $2^{[N/2]}$, where $N$
here is the number of quasiholes (``non-abelions") \cite{nayak}.

For general $SU(2)_k$ theories, we consider a $\phi^{2(k+1)}$
potential with $k+1$ minima. (The explicit potential can be written
out in terms of Chebyshev polynomials if desired.) The kinks
interpolate between adjacent vacua, so that their world lines are
domain walls in the two-dimensional picture. To count the ``number''
of such kink configurations for large $N$, one applies the same
procedure as above, and obtains $2\cos^N(\pi/(k+2))$.  In the
Read-Rezayi generalizations of the Moore-Read theory of the fractional
quantum Hall effect \cite{read-rezayi}, the statistics and the number
of states is the same \cite{slingerland}.

For the $O(3)_k$ case, the perturbation of the conformal minimal model
is different. The $\Phi_{21}$ operator is $\phi^{p-1}$, but the
resulting Landau-Ginzburg description is not as useful. However, in
section \ref{sec:braids} we showed how to describe representations of
the $SO(3)$ BMW algebra by fusing together representations of the
Temperley-Lieb algebra, and the analog is possible here. We can
describe the kinks in the Potts models as bound states of the kinks of
the $O(n)$ model, although we believe that in the $S$ matrix context
this is a formal device without physical significance. Anyway, we
consider the same potential with $k+1$ degenerate minima, but the
$O(3)_k$ kinks are comprised of two of the $SU(2)_k$ kinks. More
precisely, we consider all configurations made up of two $SU(2)_k$
kinks bound together, and then subtract off the identity.  This
procedure yields the correct fusion rules for spin-1 particles, and
the correct ``number'' of kinks $d^2-1$.

\end{document}